\documentclass[twocolumn, twocolappendix, apj]{aastex631}

\usepackage{amsmath}
\usepackage{amssymb}
\usepackage{amsthm}
\usepackage{mathtools}
\usepackage{physics}
\usepackage{bm, bbm}
\usepackage{graphicx}
\usepackage{tikz}
\usepackage{float}
\usepackage{algorithm}
\usepackage{algorithmic}
\usepackage{xcolor}

\DeclareRobustCommand{\bbone}{\text{\usefont{U}{bbold}{m}{n}1}}
\DeclareMathOperator*{\argmax}{arg\,max}
\DeclareMathOperator*{\argmin}{arg\,min}

\newtheorem{theorem}{Theorem}[section]

\newtheorem{definition}[theorem]{Definition}

\newcommand{\Sec}[1]{{\protect\hyperref[sec:#1]{Section~\ref*{sec:#1}}}}

\newcommand{\Fig}[1]{{\protect\hyperref[fig:#1]{Figure~\ref*{fig:#1}}}}

\newcommand{\subFig}[2]{{\protect\hyperref[fig:#1]{Figure~\ref*{fig:#1}~#2}}}
\newcommand{\Equ}[1]{{\protect\hyperref[equ:#1]{Equation~\ref*{equ:#1}}}}

\newcommand{\Tab}[1]{{\protect\hyperref[tab:#1]{Table~\ref*{tab:#1}}}}

\newcommand{\App}[1]{{\protect\hyperref[app:#1]{Appendix~\ref*{app:#1}}}}

\newcommand{\Alg}[1]{{\protect\hyperref[alg:#1]{Algorithm~\ref*{alg:#1}}}}

\shorttitle{Correcting Misspecified, High-Dimensional Data-Driven Priors for Inverse Problems}
\shortauthors{Barco et al.}

\begin{document}

\title{Tackling the Problem of Distributional Shifts: Correcting Misspecified, High-Dimensional Data-Driven Priors for Inverse Problems}

\author[0009-0008-5839-5937]{Gabriel Missael Barco}
\affiliation{Ciela Institute, Montréal, Canada}
\affiliation{Mila - Quebec Artificial Intelligence Institute, Montréal, Canada}
\affiliation{Department of Physics, Université de Montréal, Montréal, Canada}

\author[0000-0001-8806-7936]{Alexandre Adam}
\affiliation{Ciela Institute, Montréal, Canada}
\affiliation{Mila - Quebec Artificial Intelligence Institute, Montréal, Canada}
\affiliation{Department of Physics, Université de Montréal, Montréal, Canada}

\author[0000-0002-9086-6398]{Connor Stone}
\affiliation{Ciela Institute, Montréal, Canada}
\affiliation{Mila - Quebec Artificial Intelligence Institute, Montréal, Canada}
\affiliation{Department of Physics, Université de Montréal, Montréal, Canada}

\author[0000-0002-8669-5733]{Yashar Hezaveh}
\affiliation{Ciela Institute, Montréal, Canada}
\affiliation{Mila - Quebec Artificial Intelligence Institute, Montréal, Canada}
\affiliation{Department of Physics, Université de Montréal, Montréal, Canada}
\affiliation{Center for Computational Astrophysics, Flatiron Institute, New York, USA}
\affiliation{Perimeter Institute for Theoretical Physics, Waterloo, Canada}
\affiliation{Trottier Space Institute, McGill University, Montréal, Canada}

\author[0000-0003-3544-3939]{Laurence Perreault-Levasseur}
\affiliation{Ciela Institute, Montréal, Canada}
\affiliation{Mila - Quebec Artificial Intelligence Institute, Montréal, Canada}
\affiliation{Department of Physics, Université de Montréal, Montréal, Canada}
\affiliation{Center for Computational Astrophysics, Flatiron Institute, New York, USA}
\affiliation{Perimeter Institute for Theoretical Physics, Waterloo, Canada}
\affiliation{Trottier Space Institute, McGill University, Montréal, Canada}

\correspondingauthor{Gabriel Missael Barco}
\email{gabriel.missael.barco@umontreal.ca}

\received{XXXX}
\revised{YYYY}
\accepted{ZZZZ}
\submitjournal{The Astrophysical Journal}

\begin{abstract}
Bayesian inference for inverse problems hinges critically on the choice of priors. In the absence of specific prior information, population-level distributions can serve as effective priors for parameters of interest. 
With the advent of machine learning, the use of data-driven population-level distributions (encoded, e.g., in a trained deep neural network) as priors is emerging as an appealing alternative to simple parametric priors in a variety of inverse problems. 
However, in many astrophysical applications, it is often difficult or even impossible to acquire independent and identically distributed samples from the underlying data-generating process of interest to train these models. In these cases, corrupted data or a surrogate, e.g. a simulator, is often used to produce training samples, meaning that there is a risk of obtaining misspecified priors.
This, in turn, can bias the inferred posteriors in ways that are difficult to quantify, which limits the potential applicability of these models in real-world scenarios. 
In this work, we propose addressing this issue by iteratively updating the population-level distributions by retraining the model with posterior samples from different sets of observations, and we showcase the potential of this method on the problem of background image reconstruction in strong gravitational lensing when score-based models are used as data-driven priors. 
We show that, starting from a misspecified prior distribution, the updated distribution becomes progressively closer to the underlying population-level distribution, and the resulting posterior samples exhibit reduced bias after several updates.
\end{abstract}

\keywords{Strong gravitational lensing (1643), Bayesian statistics (1900), Hierarchical models (1925), Prior distribution (1927), Posterior distribution (1926), Sky surveys (1464), Galaxies (573), Sampling distribution (1899), Astrostatistics distributions (1884), Bayes' Theorem (1924)}

\section{Introduction}

In the era of precision science, Bayesian inference has become a cornerstone of modern statistical data analysis. It provides a mathematical framework for inferring the probability distribution of latent parameters of interest, the posterior, in the presence of noisy observations. In this paradigm, existing knowledge, encoded in a prior distribution, is updated with new, noisy information through the likelihood function, to produce updated belief over the parameters of a problem ~\citep[e.g.][]{van_de_schoot_bayesian_2021}. A key aspect of this framework is that it can be used to sequentially update the knowledge of latent parameters of interest under successive experiments or observations. In this case, the posterior for a given observation becomes the prior for the next experiment. 

In cases where there are no prior observational constraints for a given system, the only information regarding the system is the belief that it is a random sample from an underlying population.
For example, if we wish to infer the pixel values of a galaxy that has never been observed before, we use the distribution of the pixel values across all possible galaxies (i.e. the population distribution) as a prior. 
To represent population-level distributions, an option is to draw parametric priors from theoretical considerations \citep{consonni2018prior}. 
However, such theoretically motivated choices often have a simple, parametric form, and therefore lack the flexibility to express complex distributions in high dimensions \citep[e.g.][]{rupp2004bayes,little2006calibrated,gelman2008objections}. 

In those cases, an alternative approach is to use more expressive data-driven population-level priors~\citep{BDA}. In this approach, a sample of existing data representative of the population is used to learn their distribution (i.e., by using them to train a generative model) and then used as a prior in the inference of the posterior distribution for future observations. However, in many applications, it is often the case that the true population-level data-generating distribution of interest cannot be sampled directly to obtain training examples. Instead, in those cases, only corrupted (e.g. through noisy or partial observations) or approximated (e.g. simulations) samples can be obtained.

For example, for interferometric imaging, partial information about sky brightness is used to infer structures at resolutions inaccessible to single-aperture telescopes. A prime example is the Event Horizon Telescope \citep{EHT}, which is used to constrain structure at scales $\sim 25\, \mu \mathrm{arcsec}$. Learning a prior over these structures with real data of populations of black hole accretion disks would require telescopes with an aperture exceeding planet Earth's size, which is impossible. 

In cases such as those, a possible alternative is to use simulators to obtain the dataset needed to learn a population-level prior distribution. In astrophysics, high-quality simulators are often available and make this a promising avenue. For example, in cosmology, sophisticated simulations of the Universe \citep[e.g.][]{Generl2014-illustris, Rose2024-DREAMS, CMD} can provide samples of fields and objects of interest \citep[e.g.][]{Bottrell2024}. Another alternative is to construct training datasets from existing data sources that exhibit some structural similarity to the physical phenomena of interest \citep{Feng2023,Adam2023,Dia2023}. As an example, one can use high-resolution data of local, evolved galaxies to learn a prior to be applied to the analysis of high-redshift, distant, young galaxy data \citep{Smith2022, Adam2022}.

\begin{figure}[t]
    \centering
    \includegraphics[width=3.4in]{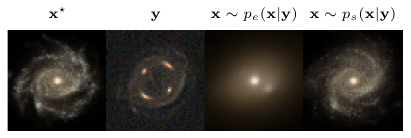}
    \caption{Impact of having a misspecified prior for posterior sampling in strong gravitational lensing source reconstruction. The first panel ($\mathbf{x}^\star$) displays the true source spiral galaxy, followed by the observed lensed image $\mathbf{y}$, which has Gaussian additive noise. Subsequent panels illustrate reconstructions using $p_e(\mathbf{x})$ (trained on elliptical galaxies, out-of-distribution) and $p_s(\mathbf{x})$ (trained on spiral galaxies, in-distribution). The reconstruction with $p_e(\mathbf{x})$ is biased due to the mismatch between the training data distribution and the distribution the observation was generated from, demonstrating the critical need for correct prior selection to ensure accurate source reconstructions.}
    \label{fig:bias_posterior}
\end{figure}

However, all these options bear the risk of a distribution shift between the learned prior and the true population-level distribution over the data-generating process of interest: even the best simulations are imperfect and only partially represent true natural processes, while noisy or similar data clearly contain features that differ from or are absent in the target distribution. As an example of the negative effect this could have, in \Fig{bias_posterior}, we compare the inference of the surface brightness of strongly lensed galaxies when adopting a misspecified prior as opposed to a correct prior.

In this work, we propose an iterative method to update an initially biased data-driven prior that approximates the process of hierarchical Bayesian inference of the population-level prior over multiple iterations of observations. We note that, while in the final stages of the preparation of this manuscript, \citet{Rozet2024} proposed a similar approach. However, instead of starting from an initial misspecified prior trained on corrupted or partial data, the initial prior proposed is a mixture of Gaussians, making the update process less efficient than what we propose here. In this work, we provide empirical results in high-dimensional settings, showing that this method can forget artifacts present in the initial prior but absent from the true data-generating distribution.  We also show that we can learn structures absent from the initial prior. This represents a viable path towards addressing the issue of distributional shifts in data-driven priors. 

To demonstrate our method, we use score-based models (SBM), as they have emerged as state-of-the-art generative models for images in recent years \citep{Ho2020ddpm,Song2020improved,Kingma2021vdm,Nichol2021improved,Dhariwal2021beatGAN,Karras2022design}. They have enjoyed a wide range of applicability in inverse problem settings due in part to the flexibility of the stochastic differential equation (SDE) formalism \citep{Song2021sde} that underpins the framework \citep{Sohl-Dickstein2015diffusion}. Examples include magnetic resonance imaging \citep[e.g.][]{Song2022mri,Chung2022mri}, super-resolution \citep[e.g.][]{Rozet2023}, deblurring, inpainting \citep[e.g.][]{Kawar2022inverse,Dou2024inverse}, phase retrieval and motion deblur \citep{Chung2023inverse}. 
In astronomy, they have been successfully applied to, among others, interferometric imaging \citep[e.g.][]{Feng2023,Dia2023,Drozdova2024,Feng2024}, deconvolution \citep[e.g.][]{Xue2023,Adam2023}, mass modelling \citep{Remy2023}, cosmological fields or cosmological parameters recovery \citep{Mudur2022,Heurtel-Depeiges2023,Floss2024,Ono2024,Mudur2024,ronan2024}, and strong gravitational lensing source reconstruction \citep{Adam2022,Karchev2022}. 

Here, we showcase our method on the problem of learning idealized brightness structures from noisy and partial observations in the inverse problem setting of strong gravitational lensing source reconstruction. The underlying motivation for this choice is that data-driven priors representing the possible population of undistorted background galaxies for this problem are often trained on either the output of hydrodynamical simulations or real local, high-resolution images of galaxies, which we expect to be morphologically different from high-redshift sources in real lenses.
In \Sec{method}, we present our method. In \Sec{results}, we present our experiments and results, and we present our discussion and conclude in \Sec{discussion}.

\section{Methods}
\label{sec:method}

\subsection{Score-based Models}

Score-based models (SBMs) are a class of generative models that aim to learn the score function of a distribution $p(\mathbf{x})$, convolved with varying levels of noise indexed by a time parameter $t$. This score function is defined as:
\begin{equation}
    \grad_{\mathbf{x}_t} \log p_t(\mathbf{x}_t) = \grad_{\mathbf{x}_t} \log \int p(\mathbf{x})p_t(\mathbf{x}_t \mid \mathbf{x})d\mathbf{x}
\end{equation}
where $p_t(\mathbf{x}_t \mid \mathbf{x})$ is the perturbation kernel that describes the noising process, and $\mathbf{x}_t$ represents the data distribution at time $t$, convolved with noise~\citep{Song2021sde}.

The perturbation kernel is generally Gaussian:
\begin{equation}
    p(\mathbf{x}_t \mid \mathbf{x}) = \mathcal{N}(\mathbf{x}_t \mid \mu(t)\mathbf{x}, \sigma^2(t)\bbone),
\end{equation}
where $\mu (t)$ and $\sigma^2(t)$ define the mean and variance of the noise. Different choices of these functions result in different SDEs. In this work, we use two types of SDEs: the Variance Exploding (VE) SDE \citep{Song2019} and the Variance Preserving (VP) SDE \citep{Ho2020ddpm, Song2021sde}. We explain the choice of either VP or VE in each case in \App{model_architecture}. The VE SDE is defined by
\begin{equation}
    \mu(t) = \bbone \quad , \quad
    \sigma(t) = \sigma_{\text{min}} \left(\frac{\sigma_{\text{max}}}{ \sigma_{\text{min}}}\right)^{t},
\end{equation}
while the VP SDE is characterized by
\begin{align}
    \mu(t) & = e^{-\frac{1}{4}t^2(\beta_{\text{max}} - \beta_{\text{min}}) - \frac{1}{2}t \beta_{\text{min}}} \\
    \sigma(t) & = \sqrt{1 - e^{-\frac{1}{2}t^2(\beta_{\text{max}} - \beta_{\text{min}}) - t \beta_{\text{min}}}}.
\end{align}

A neural network, $\mathbf{s}_\theta(\mathbf{x}, t)$, typically a U-net \citep{Ronnenberger2015}, is trained to approximate $\grad_{\mathbf{x}} \log p_t(\mathbf{x})$ using independent and identically distributed ($\mathrm{iid}$) samples $\mathbf{x} \in \mathbb{R}^n$ from the data $\mathcal{D} \overset{\mathrm{iid}}{\sim} p(\mathbf{x})$. The network is optimized by minimizing the denoising score-matching objective \citep{Hyvarinen2005,Vincent2011}
\begin{equation}
\label{equ:loss}
\mathcal{L}_\theta = 
\underset{\substack{\mathbf{x} \sim \mathcal{D} \\ t \sim \mathcal{U}(0, 1) \\ \mathbf{x}_t \sim p(\mathbf{x}_t \mid  \mathbf{x})}}{\mathbb{E}}
\bigg[ \lambda(t) \big\lVert \mathbf{s}_\theta(\mathbf{x}_t, t) - \grad_{\mathbf{x}_t} \log p_t(\mathbf{x}_t \mid \mathbf{x})  \big\rVert^2 \bigg]\, ,
\end{equation}
where the expectation is taken over samples from the distribution $\mathbf{x} \sim \mathcal{D}$, different time steps $t \sim \mathcal{U}(0, 1)$, and conditional noisy samples from $\mathbf{x}_t \sim p_t(\mathbf{x}_t \mid \mathbf{x})$.

$\lambda(t)$ is a suitably chosen weight. We use $\lambda(t) = \sigma^2(t)$ following the work of \citet{Song2021sde}. In particular, for VE, this choice corresponds to approximate maximum likelihood training~\citep{Song2021MLE}

We use the \texttt{score-models}\footnote{\href{https://github.com/AlexandreAdam/score\_models}{github.com/AlexandreAdam/score\_models}} package for creating and training the models $\mathbf{s}_\theta(\mathbf{x}, t)$. The package provides pre-configured architectures and optimized training routines for SBMs, allowing efficient model training. Details on the model architecture and hyperparameters used in this work are provided in the \App{model_architecture}.

Having access to an approximation of the score function allows one to create a generative model by solving the reverse-time SDE \citep{Anderson1982} associated with the noising process used during training:
\begin{equation}
\label{equ:sde}
    d \mathbf{x} = (f(\mathbf{x}, t) - g^2(t) \grad_{\mathbf{x}} \log p_t(\mathbf{x})) dt 
    + g(t) d \bar{\mathbf{w}}\, ,
\end{equation}
where $f$ is the drift, $g$ is an homogeneous diffusion coefficient associated with the noising process, and $\bar{\mathbf{w}}$ is a reverse-time Wiener process. 
This generative process can also be used for inference by conditioning the probability on new data, thus using the posterior score function, $\grad_\mathbf{x} \log p(\mathbf{x} \mid \mathbf{y})$, in the reverse-time SDE \eqref{equ:sde}. 
In the next section, we describe the linear inverse problem setting in which this framework is applied.

\subsection{Score-based Priors for Linear Inverse Problems}

\label{sec:inverse}
 
A linear inverse problem is characterized by the equation
\begin{equation}
 \mathbf{y} = A\mathbf{x} + \boldsymbol{\eta}
\end{equation}
where $\mathbf{x} \in \mathbb{R}^{n}$ are the parameters of interest, $\mathbf{y} \in \mathbb{R}^m$ is the observation, and $\boldsymbol{\eta} \in \mathbb{R}^m$ is a vector of additive noise. For a linear inverse problem, the observation and parameters of interests are related by a constant matrix $A \in \mathbb{R}^{m \times n}$.

In Bayesian inference, the goal is to sample from the posterior distribution, $ p(\mathbf{x} \mid \mathbf{y})$. This can be accomplished by replacing the prior score function in the reverse-time SDE \eqref{equ:sde} with the posterior score function, which is obtained using Bayes' theorem.
However, the likelihood score is intractable, as it involves an expectation over backward trajectories of the reverse-time SDE \citep[see e.g.][]{Feng2023}. 
For a Gaussian likelihood, we can construct an analytical estimate of its score using the convolved likelihood approximation \citep{Remy2023, Adam2022}
\begin{equation}
    p_t(\mathbf{y} \mid \mathbf{x})\approx \mathcal{N}(\mu(t)\mathbf{y} \mid  A\mathbf{x}, \mu^2(t)\Sigma_{\boldsymbol{\eta}} + \sigma^2(t) AA^T)\, ,
\end{equation}
where $\Sigma_{\boldsymbol{\eta}}$ is the covariance associated with the additive noise distribution $\boldsymbol{\eta} \sim \mathcal{N}(0, \Sigma_{\boldsymbol{\eta}})$. 
With this machinery, any SBM trained on some dataset of parameters of interests, $\mathcal{D} \overset{\mathrm{iid}}{\sim} p(\mathbf{x})$, can be used as an approximate posterior sampler in zero shots --- i.e. without retraining or conditioning the neural network on the observations \citep{Graikos2022_plug_and_play}.

\subsection{Updating the Prior from Observations}
\label{sec:prior_updates}

\begin{figure}[tb]
    \centering
    \begin{tikzpicture}
        \tikzset{arrow style/.style={->, thick, >=latex}}
        \node[draw, thick, circle, minimum size=0.7cm] (eta) at (0, 1.2) {$\boldsymbol{\eta}_i$};
        \node (theta) at (4.2, 1.2) {$\theta^\star$};
        \node[draw, thick, circle, minimum size=0.7cm] (x) at (3, 1.2) {$\mathbf{x}_i$};
        \node[draw, thick, circle, fill=gray!30, minimum size=0.7cm] (y) at (1.5, 0.5) {$\mathbf{y}_i$};
        \draw[arrow style] (eta) -- (y);
        \draw[arrow style] (theta) -- (x);
        \draw[arrow style] (x) -- (y) node[pos=0.3, below] {$A$};
        \node at (2.6, -0.2) {$i = 1, \ldots, N$};
        \draw[solid, thick, rounded corners] (-0.7, -0.5) rectangle (3.7, 1.85);
    \end{tikzpicture}
    \caption{Graphical model of the inference problem. The \textit{true} prior distribution, parametrized by the population-level parameters $\theta^\star$, generates the realizations of unobserved parameters of interests $\mathbf{x}_i$. Our goal is to learn an estimate $\hat{\theta} \approx \theta^\star$. In this work, we have access to the noise distribution that generates $\boldsymbol{\eta}_i$, the forward model $A$ and a set of $N$ observations $\{\mathbf{y}_i\}_{i=1}^N$.}
    \label{fig:graphical_model}
\end{figure}
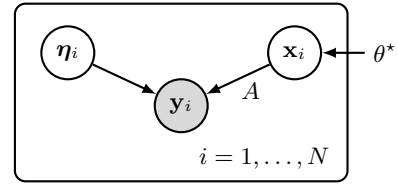

We aim to study hierarchical inference in the context of moderately high-dimensional inference inverse problems using SBM as expressive priors. We illustrate the data-generation process for this problem in \Fig{graphical_model}, where $p_{\theta^\star}$ is the true population-level distribution describing the underlying prior distribution we wish to approximate. 
Assuming an initial SBM prior trained on a potentially corrupted dataset $\{\mathbf{x}^{(0)}_i\}$, our goal is to update the population-level parameters $\theta$ --- the parameters, or weights, of the prior SBM network --- given only a set of noisy/partial observations $\{\mathbf{y}_i\}_{i=1}^N$. In a Bayesian framework, this amounts to calculating:
\begin{equation}
    p(\theta \mid \{\mathbf{y}_i\}_{i=1}^N)=\frac{p(\theta)}{\mathcal{Z}} \prod\limits_i \int p(\mathbf{y}_i \mid \mathbf{x})p(\mathbf{x} \mid \theta)d\mathbf{x}\,.
\end{equation}
However, this approach is not practical in large dimensional spaces since using a distribution over $\theta$ would amount to consider distributions over SBMs as a hyper-prior to perform inference, which would be intractable. A more feasible approach, which we will adopt here, is to approximate the distribution over $\theta$ by its maximum marginal likelihood point estimate using empirical Bayes methods~\citep{introEmpiricalBayes}:
\begin{equation}
\label{eqn:thetastar}
    \hat\theta = \argmax_{\theta \in \Theta}\, p(\{\mathbf{y}_i\}_{i = 1}^N\mid\theta) \, .
\end{equation}
However, even this simplified setting poses a challenging task for diffusion models since the loss function \eqref{equ:loss} of an SBM cannot be easily modified to relate parameters from the target distribution, $\mathbf{x}$, with the observation, $\mathbf{y}$.
This is unlike other generative models like VAE \citep{Kingma2014} or Normalizing Flows \citep{Dinh2014, Rezende2015_NF} which have inexpensive sampling procedures and sufficient inductive biases to allow implicit training through an explicit transformation between an abstract, lower-dimensional latent space
and observation space \citep[e.g.][]{lanusse_deep_2021,Sun2020}. 

To circumvent this issue, we introduce a method inspired by traditional generalized expectation maximization methods~\citep[e.g.][]{Hartley1958,Dempster1977,mclachlan2007algorithm, RuthWilliam2024,Rozet2024}. We note that source distribution estimation techniques \citep[e.g.][]{vandegar2021} have also been developed to deal with a similar problem. However, to our knowledge, these techniques have only been applied to low-dimensional settings.

Our method consists of an iterative procedure that leverages the posterior sampling algorithm outlined in section \ref{sec:inverse} to acquire increasingly plausible samples from the set of observations $\{\mathbf{y}_i^{(\alpha)} \}_{i=1}^N$.
For each update, labeled by $\alpha \in \{1,\dots,M\}$, a set of $K$ posterior samples, $\{\mathbf{x}_{i,j}^{(\alpha)}\}_{j=1}^K$, is aggregated from each observation to train a new prior distribution, $p_{\theta_{\alpha + 1}}(\mathbf{x})$, encoded by the generative process of an SBM, $\mathbf{s}_{\theta_{\alpha + 1}}(\mathbf{x}, t)$.
More precisely, at each iteration, we wish to train an SBM with parameters $\theta_{\alpha+1}$ that maximizes the log-likelihood of posterior samples obtained for a set of observations using the previous prior $p_{\theta_\alpha}(\mathbf{x})$. That is, we want to find $\theta_{\alpha + 1}$ such that the updated prior given by $\mathbf{s}_{\theta_{\alpha + 1}}(\mathbf{x}, t)$ approximates $\mathbb{E}_{\mathbf{y}\sim p(\mathbf{y})} p_{\theta_\alpha}(\mathbf{x} \mid \mathbf{y})$ in the large data limit.
This is equivalent to finding the set of $\theta_{\alpha+1}$ minimizing the KL divergence~\citep{KLdiv}:
\begin{eqnarray}
\theta_{\alpha+1}&=&\displaystyle\argmin_{\theta \in \Theta} \textrm{KL}\left( \int d\mathbf{y}\, p(\mathbf{y}) p_{{\theta_\alpha}} (\mathbf{x} \mid \mathbf{y})\, \Big\Vert\, p_{\theta} (\mathbf{x})\right)\nonumber
\end{eqnarray}
\begin{eqnarray}
    &=& \displaystyle\argmax_{\theta \in \Theta} \int d\mathbf{y} d\mathbf{x}\, p(\mathbf{y}) p_{{\theta_\alpha}} (\mathbf{x} \mid \mathbf{y}) \log  p_{\theta} (\mathbf{x})\nonumber\\
    &=&  \displaystyle\argmax_{\theta \in \Theta} \underset{\substack{\mathbf{y}\sim p(\mathbf{y}) \\ \mathbf{x}\sim p_{{\theta_\alpha}} (\mathbf{x} \mid \mathbf{y})}}{\mathbb{E}}\left[ \log  p_{\theta} (\mathbf{x})\right]
\end{eqnarray}

\begin{definition}
\label{def:mixture_prior}
Given a prior $p_{\theta_\alpha}(\mathbf{x})$ and a set of observations $\mathcal{S} = \{\mathbf{y}_i^{(\alpha)}\}_{i = 1}^N$, we define the next prior distribution $p_{\theta_{\alpha + 1}}(\mathbf{x})$ as the distribution encoded by the generative process of the SBM $\mathbf{s}_{\theta_{\alpha + 1}}(\mathbf{x}, t)$ trained by minimizing the denoising objective (\ref{equ:loss}) with training set $\mathcal{D}$ given by
\begin{equation}
    \mathcal{D}=\{\mathbf{x}_{i,j} \mid \mathbf{x}_{i,j} \sim p_{\theta_\alpha} (\mathbf{x}\mid \mathbf{y}_j) ,\, \forall i \in [1,K] ,\, \forall \mathbf{y}_j\in \mathcal{S} \}
\end{equation}

\end{definition}

Our algorithm is summarized in \Alg{alg}. Since the population parameters $\theta_\alpha$ are the parameters of an over-parametrized neural network, the sequence $\theta_0,\dots, \theta_M$ is not guaranteed to converge to the unique true distribution represented by $\theta^\star$.
However, we can still make statements about the convergence of the sequence of priors $p_{\theta_0}(\mathbf{x}), \dots, p_{\theta_M}(\mathbf{x})$ to a local extremum of the evidence which will satisfy Eqn.~(\ref{eqn:thetastar}). 
We can show that the updated prior will have larger log-evidence than the previous iteration (see \App{proof}). Furthermore, in the large data limit, $N \rightarrow \infty$, we can show that there exists a stationary representation for the prior distribution, $p_{\hat\theta}(\mathbf{x})$, which has log-evidence equal to the true distribution represented by $\theta^\star$ (see \App{proof_stationary}).

We show empirically in sec.~\ref{sec:results} that the procedure does converge to a prior close to the true underlying population distribution for the settings we explored.
A similar phenomenon has been observed by \citet{Kadkodaie2023}, where 
diffusion models trained on different subsets of a dataset converge to the same distribution, as well as to the same score function representation, $s_{\theta}(\mathbf{x})$, when trained in the large data limit. 
Their work can be viewed as a special case of our method where $M=1$, $A = \bbone$, and $\Sigma_{\boldsymbol{\eta}} = 0$.

In our experiment, we repeat the update \ref{def:mixture_prior} iteratively by gathering a new set of observations at each iteration $\alpha$. The core assumption in the construction of the \Alg{alg} is that the number of observations available to construct a dataset to train an SBM at each iteration is sufficiently large to be considered in the large data limit. In this work, we use simulators to satisfy this criterion. In practice, specifically in the case of strong gravitational lensing data, upcoming large sky surveys like the Legacy Survey of Space and Time (LSST) at the Rubin Observatory and the Euclid space telescope are expected to discover of order 200,000 new strong lenses~\citep{Collett2015}, and we will therefore easily have access to sufficiently large datasets to afford updates such as the scheme proposed here. 

\begin{algorithm}[H]
   \caption{Updating the Prior with Observations}
   \label{alg:alg}
\begin{algorithmic}
   \STATE {\bfseries Input:} Initial prior $p_{\theta_0}(\mathbf{x})$, observations $\{\mathbf{y}\}_{i=1}^{M \times N}$, training procedure $\mathcal{A}$, posterior sampling procedure $\mathcal{F}$.
   \FOR{$\alpha=1$ {\bfseries to} $M$}
   \STATE Select $N$ observations $\{\mathbf{y}_i^{(\alpha)}\}_{i = 1}^N = \{\mathbf{y}_i\}_{i = (\alpha -1)N}^{\alpha N}$
   \FOR{$i=1$ {\bfseries to} $N$}
   \STATE Get $K$ posterior samples $\mathbf{x}_{i, j}^{(\alpha)} \sim p_{\theta_{\alpha-1}}(\mathbf{x} \mid \mathbf{y}_i^{(\alpha)})$
   \STATE $\mathcal{D}_{\text{posterior}}^{i, \alpha} = \{\mathbf{x}_{i, j}^{(\alpha)}\}_{j=1}^K = \mathcal{F}(\mathbf{y}_i^{(\alpha)}, p_{\theta_{\alpha-1}}(\mathbf{x}), K)$
   \ENDFOR
   \STATE Create proposal dataset $\mathcal{D}_{\text{proposal}} = \bigcup_{i = 1}^N \mathcal{D}_{\text{posterior}}^{i, \alpha}$
   \STATE Train new prior $p_{\theta_{\alpha}} = \mathcal{A}(\mathcal{D}_{\text{proposal}})$
   \ENDFOR
\end{algorithmic}
\end{algorithm}

\section{Forward model: Strong Gravitational Lensing}
\label{sec:forward_model}

Strong gravitational lensing occurs when a massive object curves space in such a way that the light emitted by a distant background source, for example, a galaxy, is deflected, causing the image we see from the source to be distorted and multiply imaged. Given a noisy lensed observation, reconstructing the source galaxy and the mass distribution in the lens is an important scientific problem, which can act as a probe to understand the nature of dark matter~\citep[e.g.][]{Hezaveh2016,Vegetti2014}, the early formation of stars~\citep[e.g.][]{Welch2022}, active galactic nuclei~\citep[e.g.][]{Peng2006}, the expansion rate of the universe~\citep[e.g.][]{holycow}, and other related problems.

In the limit of a thin lens, and assuming that the mass distribution in the foreground lens is known, strong gravitational lensing of a background source image into a distorted observation is a linear transformation. To simulate the forward strong gravitational lensing model, we use the \texttt{Caustics} python package \citep{Stone2024}. 

In our experiments, we employ a Singular Isothermal Ellipsoid (SIE) lens model within a Flat-$\Lambda$CDM cosmology. The background source is represented by a grid of pixels. For simplicity, we fix the ellipticity of the lens model to $q = 0.5$, and the position angle to $\phi = \pi /5$. The values for the Einstein radius, $\theta_E$, and the pixel scale of the source depend on the experiments and are reported in section \ref{sec:results}. The other parameters are set to the \texttt{Caustics} default values.

With these parameters defined, we can obtain the transformation matrix $A$ by computing the Jacobian matrix of the simulator. 
Finally, we add Gaussian noise to the simulations.

\section{Experiments and Results}

\label{sec:results}

We conduct experiments with galaxy images and with MNIST \citep{LeCun1998-mnist}, a dataset of handwritten digits with a $28\times 28$ resolution. For the former, we gather what we refer to as an \textit{elliptical dataset} and a \textit{spiral dataset}, which are detailed in the next section.

For both sets of experiments, we intentionally create a misspecified initial prior $p_{\theta_0}(\mathbf{x})$ compared to the true underlying distribution $p_{\theta^\star}(\mathbf{x})$, either by dropping modes --- i.e. digits in the MNIST dataset --- or by training on different types of objects --- spiral or elliptical galaxies. 
In our experiments, $p_{\theta^\star}(\mathbf{x})$ is defined by another SBM. The hyperparameters for all the SBMs trained in this work are reported in \App{model_architecture}.
We also assume that the forward model, characterized by the lensing matrix $A$ and described in \Sec{forward_model}, is the same for all observations and is known to the observer. For MNIST we set $\theta_E = 0.5''$ and the source pixel scale to $0.03''$, while for galaxies, these constants are set to $ 0.8''$ and $0.04''$, respectively.

Finally, the number of iterations chosen for each experiment varies. For MNIST, we performed only $M = 4$ prior updates, as this serves as a proof-of-concept example, and the correct digits are already inferred accurately using the final prior $p_{\theta_4}(\mathbf{x})$. However, the class proportions could potentially be further improved with additional iterations. In contrast, for the galaxy experiment, we carried out $M = 10$ iterations, stopping when convergence was achieved based on the metrics discussed in subsequent sections.

\subsection{MNIST: Mode Mismatch}
\label{sec:mnist_exp}

As a first experiment, we explore mode mismatch, which we define as the situation where the initial prior and the true underlying population-level distribution do not share the same modes. As an example, we train the initial prior $p_{\theta_0}(\mathbf{x})$ on a subset of MNIST with the digits 1 and 4 removed, while the true population distribution, $p_{\theta^\star}(\mathbf{x})$, is constructed with the digits 1 and 6 missing.

\begin{figure}[tb]
    \centering
    \includegraphics[width = 3.3in]{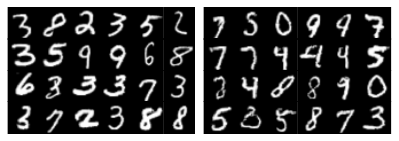}
    \caption{This figure demonstrates, with prior samples, the effectiveness of model updates in learning and forgetting specific digits in the MNIST experiment under observational noise of $\sigma_{\boldsymbol{\eta}} = 0.5$. \textbf{Left}: Samples of initial model $p_{\theta_0}$. These samples include the digit $6$ and lack the digit $4$, in agreement with the initial model's training distribution. \textbf{Right:} Samples from the final prior $p_{\theta_4}$. In this model, the digit $6$ is absent, and digits resembling $4$ are present, showcasing successful adaptation to the target distribution. However, sample quality is variable, as seen in the top left sample, where it is ambiguous whether it represents a $1$ or a $7$.}
    \label{fig:bad_samples}
\end{figure}

We perform $M = 4$ iterations using our algorithm. Samples from the initial distribution and the final prior update are shown in \Fig{bad_samples}, demonstrating the model's effectiveness in learning and forgetting specific digits. However, we also observe a decline in the quality of prior samples $\mathbf{x} \sim p_{\theta_4}(\mathbf{x})$, which could be attributed to the approximations made during posterior sampling $\mathbf{x} \sim p(\mathbf{x} \mid \mathbf{y})$ or the iterative retraining with generated samples \citep[see e.g.][]{bertrand2024on}.

In each update, $N = 60\,000$ observations are generated and $K=1$ posterior samples are generated for each of them to train the next SBM prior. We use a Predictor-Corrector SDE solver with $500$ steps ($3$ Langevin corrector step per predictor step) as posterior sampling procedure $\mathcal{F}$.

\begin{figure*}[tb]
    \centering
    \includegraphics[width = 7.1in]{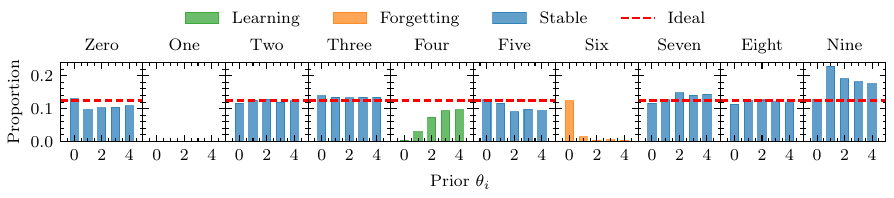}
    \caption{Learning and forgetting dynamics across updates using \Alg{alg} towards the target distribution in the MNIST experiment with observational noise of $\sigma_{\boldsymbol{\eta}} = 0.4$ during the updates. The plot shows the classification of $2\,048$ prior samples $\mathbf{x} \sim p_{\theta_\alpha}(\mathbf{x})$ at each update by a CNN classifier. Each panel corresponds to a digit category, with proportions of samples being shown for each update. Initially, the prior was trained excluding digits $4$ and $1$, while the target distribution excluded $6$ and $1$. The objective was to forget digit $6$, which was accomplished in a single update, and to learn digit $4$, which took several updates, initially producing classifications similar to digit $9$. The red dashed line represents the proportion from the target distribution, serving as a benchmark.}
    \label{fig:mnist_number_counts}
\end{figure*}

To systematically identify the digits, we train a CNN classifier on MNIST. This allows us to track the proportion of each mode in the prior distributions as a function of the prior update. These results are shown in \Fig{mnist_number_counts}. We observe that the digit $6$ is dropped after the first iteration. This is expected since our algorithm only uses posterior samples for its update, and no observations were consistent with the number 6 in the first round given the chosen inverse problem and noise level. On the other hand, learning to infer the digit $4$ requires multiple iterations. Interestingly, the proportion of $9$'s correspondingly increases after the first iteration (as the digits $4$ and $9$ are morphologically very similar, and therefore true $4$s get reconstructed as $9$s by the initially misspecified prior) and gradually decreases as the proportion stabilizes closer to the true data-generating distribution. All other numbers keep almost the same proportion. 

\begin{figure}[tb]
    \centering
    \includegraphics[width = 3.4in]{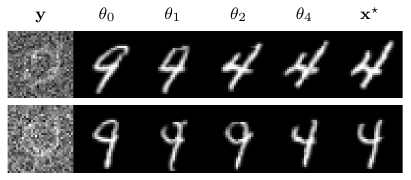}
    \caption{Sequences showcasing the model's ability to accurately reconstruct the digit $4$ from noisy observations in posterior samples after updating the prior with \Alg{alg}, despite $4$ not being included in the initial prior. The evolution of posterior samples $\mathbf{x} \sim p_{\theta_\alpha}(\mathbf{x} \mid \mathbf{y})$ is shown for each update, highlighting the model's improvement. Starting from the observed image $\mathbf{y}$ (leftmost), the samples progress from $\theta_0$ to $\theta_4$ (left to right). Initially, the model confuses $4$ with $9$ under $\theta_0$ and $\theta_1$. By $\theta_2$ and, more clearly, by $\theta_4$, the posterior accurately reflects $4$, closely matching the true digit $\mathbf{x}^\star$ (rightmost). This illustrates the effectiveness of the model in adapting its posterior estimates over sequential updates to learn and correct its representation of digit $4$ from noisy data.}
    \label{fig:mnist_posterior}
\end{figure}

The initial bias in inferred digits $9$ which we hypothesize to be due to the similarity between the digits $4$ and $9$ can be intuitively visualized in \Fig{mnist_posterior}, where samples from the posterior distribution of each update are shown for $2$ observations generated from the number $4$. In the first few columns, the posterior samples are biased toward reconstructing the digit $9$ because $4$ is missing from this prior. Crucially, our algorithm can recover the correct shape of the digit 4 after a few updates, even though this digit was never seen during the initial training of the prior SBM. The accuracy of the learnt prior can be quantified with further metrics as presented in \App{details_mnist}.

\subsection{Galaxies: Distribution Shift in High-dimensional Space}
\label{sec:galaxies}

We also test the proposed method in a more realistic setting, with the scientific application of recovering undistorted images of strongly lensed galaxies. To this end, we gather two datasets of galaxies of different classes: simulated observations of blue spiral galaxies and real observations of red elliptical galaxies as described in ~\Sec{data}. This choice is motivated by the fact that initially, only noisy telescope observation of local, evolved red galaxies would be available to train an initial prior, but we want to demonstrate that the proposed method can discover new features in the data (here, spiral arms), and recover the underlying, noise-free population-level prior, which we represent as noise-free blue galaxy obtained through simulations.

We define the true prior distribution $p_{\theta^\star}$ to be what an SBM learned when trained on the spiral dataset. 
The initial mismatched distribution $p_{\theta_0}$ is an SBM trained on the elliptical dataset. 
Both these models and the ones we get during the updates have the same architecture and training hyperparameters. 

\subsubsection{Data}
\label{sec:data}

\begin{figure}[tb]
    \centering
    \includegraphics[width = 3.4in]{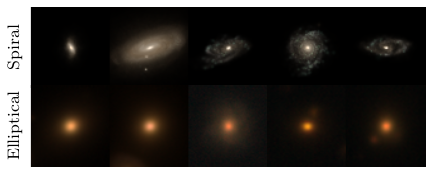}
    \caption{Random samples from the two galaxy datasets used in this work, highlighting the distinction between both. The \textit{spiral galaxy dataset}, featuring five samples in the top row, comes from a subset of the SKIRT TNG dataset \citep{Bottrell2024} and showcases a range of morphological features typical of this galaxy class. The \textit{elliptical galaxy dataset}, presented in the bottom row with five samples, is sourced from the DESI Legacy Imaging Surveys \citep{Dey_2019} and is characterized by smooth, featureless light profiles without significant structural details. Further details about both datasets are available in \Sec{data}.}
    \label{fig:data}
\end{figure}

The \textbf{spiral dataset} is a synthetic dataset used as the true distribution for our experiments. It is taken from the SKIRT TNG dataset \citep{Bottrell2024}, made by a large public collection of images covering bands from $0.3$-$5$ microns made by applying dust radiative transfer post-processing \citep{Camps2020} to galaxies from the TNG cosmological magneto-hydrodynamical simulations\footnote{\url{www.tng-project.org}} \citep{Nelson2019}. 
This synthetic data is simulated for the \textit{grz} filters of the Hyper Suprime-Cam Subaru Strategic Program \citep[][]{HSCSurvey} and assigned to the $(B, G, R)$ color channels, respectively, and serves as our ground truth sample since it contains no observational noise and can be taken at high resolution.
We take $10\,000$ data points from this dataset, convert to flux in $\mu\mathrm{Jy}\, \mathrm{sr}^{-1}$ units, and downsample to $64\times64$ pixel images to train an SBM.

The \textbf{elliptical dataset} is used as the initial prior for our experiments. It is strongly out of distribution compared to the spiral dataset as it includes some corruption effects from real observation (e.g. observational noise and psf blurring) and is overall void of high-frequency features, unlike spiral galaxies. Moreover, the color channels are markedly different between the two sets (see \Fig{data}).
We collected $10\,459$ galaxy images from the DESI Legacy Imaging Surveys \citep{Dey_2019} DR10, selected using the SDSS-IV \citep{Blanton2017-SDSS-IV} DR17 \citep{SDSS-DR17} database via Astroquery \citep{Ginsburg2019-Astroquery} to construct this dataset. 
We selected these galaxies based on the elliptical class from GalaxyZoo \citep{Lintott2011-GalaxyZoo}, using a threshold of at least 10 votes and a probability of at least $70\%$. 
We also filter postage stamps with thresholds for total magnitude ($5 \leq$ \texttt{modelMag\_r} $\leq 22$), radius ($2^\prime \leq r \leq 20^\prime$), and flux criteria. 
Here we also select the \textit{grz} bands for this dataset and assign them to the $(B, G, R)$ color channels. 
The images are sampled at $64\times64$ pixel resolution, and the galaxy sample has been chosen to fit well in this size at the native resolution for the DESI observations.

Random samples from both datasets are shown in \Fig{data}. 

\subsubsection{Galaxy Experiment Details and Results.}

For this experiment, we consider $M=10$ updates using $N=10\,240$ observations per update, simulated by sampling $\mathbf{x} \sim p_{\theta^\star}$, $\boldsymbol{\eta} \sim  \mathcal{N}(0, \sigma_{\boldsymbol{\eta}}^2 \bbone)$, and forward modelling $\mathbf{y} = A\mathbf{x} + \boldsymbol{\eta}$. We get $K = 1$ posterior samples per observation, so the proposal datasets at each update have $10\,240$ elements. The posterior sampling procedure $\mathcal{F}$ is a Predictor-Corrector SDE solver \citep{Song2021sde} with $2\,048$ steps ($1$ Langevin corrector step per predictor step). For the updates themselves, we choose a relatively low noise level, $\sigma_{\boldsymbol{\eta}} = 0.5$, to speed up convergence. We also perform an extra experiment with $\sigma_{\boldsymbol{\eta}} = 1$.

\begin{figure*}[tb]
    \centering
    \includegraphics[width=7.1in]{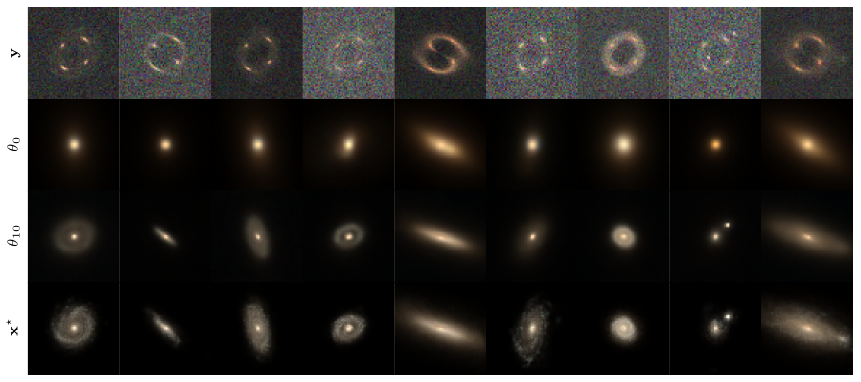}
    \caption{Improvement in strong gravitational lensing source reconstruction with galaxy sources under \Alg{alg} after $M = 10$ updates, highlighting the adaptation from a biased initial prior to better alignment with the target distribution. The top row shows noisy observations $\mathbf{y}$ with observational noise of $\sigma_{\boldsymbol{\eta}} = 3$, chosen to be high enough to showcase the distribution shift from the biased prior. The second row displays posterior samples from the initial prior $p_{\theta_0}(\mathbf{x})$, characterized by significant bias. The third row presents samples after the final update $p_{\theta_{10}}(\mathbf{x})$, demonstrating substantial improvements in matching the true sources, as represented by the bottom row ($\mathbf{x}^\star$). This sequence illustrates the algorithm's capability to refine predictions and effectively overcome initial biases.
    }
    \label{fig:main_plot}
\end{figure*}

Unless stated otherwise, for visualization purposes, galaxy images $\mathbf{x}$ are plotted with $\mathbf{\bar{x}} = \log(\mathbf{x} - \mathbf{x}.\text{min()} +1 )$, and normalized between $0$ and $1$. Training, inference, and evaluation are performed without this transformation. Visualizations of images of lensed observations $\mathbf{y}$ are only normalized between $0$ and $1$.

In \Fig{main_plot}, we show a selection of 10 observations and their posterior samples for the first and last iterations. This figure demonstrates how the structure of generated posterior samples evolves as a function of the update, increasing in complexity from elliptical shapes to showcasing rings and satellites in the last update.

Moreover, these features recover aspects of the ground truths that were completely missed when using the misspecified prior. We emphasize that the samples obtained using the initial prior are strongly biased toward elliptical galaxies and differ from the ground truths $\mathbf{x}^\star$ in color, flux, and morphology. Thus, important morphological information about the data is learned and encoded in the updated SBM which only had access to noisy observation.

Although learning of high-frequency modes can be observed in this experiment, the speed of convergence is considerably slower than that of MNIST. This might be due to the dimensionality of the problem, but also the specific configurations of this experiment. Additionally, as mentioned before, unlike the spiral dataset, the elliptical dataset generally lacks high-frequency features. Recovering these high-frequency features poses challenges, particularly when the noise level is higher ($\sigma_{\boldsymbol{\eta}} = 1$ vs. $\sigma_{\boldsymbol{\eta}} = 0.5$). This is not only because this information is dimmer and more degraded by Gaussian noise, but also because the initial prior is biased towards not having them, making it more challenging to recover. We performed a backward experiment, transitioning from an initial spiral prior to a target elliptical prior (see \App{reverse_experiment}), and although the signal-to-noise (SNR) ratio is lower, the reconstructions are good after only $6$ updates.

We hypothesize that the choice of initial prior $p_{\theta_0}(\mathbf{x})$ plays an important role in determining the distribution reached at convergence after several iterations, given a certain noise level, as well as the kind of features correctly recovered. Further exploration of this procedure with different initial priors, as well as a more thorough examination of our algorithm's rate of convergence as a function of experiment design, is left for future work.

\subsubsection{Quantification of the Convergence}

\begin{figure}[tb]
    \centering
    \includegraphics[width = 3.3in]{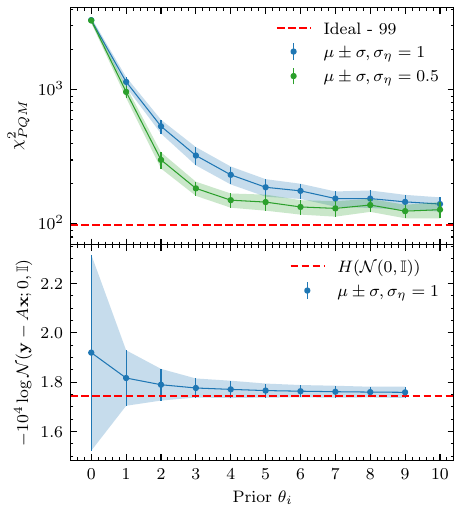}
    \caption{Evaluation of prior improvement using two statistical metrics over ten prior updates for the galaxy experiment. \textbf{Top:} The $\chi^2_{PQM}$ statistic compares the target distribution $p_{\theta^\star}$ with the proposal prior $p_{\theta_i}$ at each iteration, using samples from both distributions. The progress of the $\chi^2_{PQM}$ statistic across iterations highlights the improvement of the proposal prior towards the target prior distribution, as well as how the final values depend on the observational noise levels, $\sigma_{\boldsymbol{\eta}} = 0.5$ and $\sigma_{\boldsymbol{\eta}} = 1$. Lower $\chi^2_{PQM}$ values indicate closer alignment with the ideal scenario (dashed red line at 99). \textbf{Bottom:} Mean log-likelihood of the residuals $\mathbf{y} - A\mathbf{x}$, with data points representing mean values from $10\,240$ pairs $( \mathbf{y}, \mathbf{x} \sim p_{\theta_\alpha}(\mathbf{x} \mid \mathbf{y}))$. The red dashed line denotes the entropy of the noise distribution, serving as the target optimal value. Together, these plots demonstrate how iterative modifications in $\theta_i$ refine the prior towards the ground truth and enhance the reliability of posterior sampling for source reconstruction.}
    \label{fig:mean_llk}
\end{figure}

In \Fig{mean_llk} (bottom), we compute the log-likelihood of the residuals as a function of the update index and compute the mean and variance of the log-likelihood. The ideal value corresponds to the entropy of the noise model:
\begin{equation}
    H(\mathcal{N}(0, \Sigma_{\boldsymbol{\eta}})) = - \mathbb{E}_{\mathbf{x}\sim \mathcal{N}(0, \Sigma_{\boldsymbol{\eta}})}[\log \mathcal{N}(\mathbf{x}; 0, \Sigma_{\boldsymbol{\eta}})]
\end{equation}
This metric informs us about the information left in the data to be extracted by the posterior sampling algorithm. We observe that the mean converges to the ideal scenario while the variance reduces accordingly.

As a test that the updated prior SBM learns the correct distribution, we also compute the PQMass metric between each updated prior $p_{\theta_\alpha}(\mathbf{x})$ and $p_{\theta^\star}(\mathbf{x})$ for the galaxy experiment. 

PQMass is a sample-based metric to assess the quality of generative models \citep{lemos2024-pqmass} which is based on partitions of the space by Voronoi cells to estimate the probability that both samples come from the same distribution using a $\chi^2_{PQM}$ using counts in those cells.

We use $n_r = 100$ regions, which corresponds to $n_r -1 = 99$ degrees of freedom. We estimate a mean and variance by using 5 independent sets of samples from both distributions, $p_{\theta^\star}$ and the proposal prior distribution $p_{\theta_i}$, with $2,048$ samples in each set. Then, for each pair of sets, we re-compute the value with different regions $10$ times. The results are reported in the top panel of \Fig{mean_llk}.

We observed that the value of $\chi^2_{PQM}$ improves as a function of the update index. The experiment with lower noise levels improves faster to a final value of $\chi^2_{PQM} = 127.87\pm 17.86$, while the higher noise level experiment reaches only $\chi^2_{PQM} = 140.93\pm 17.08$.

\subsubsection{Axis Ratio}\label{sec:axisratio}

\begin{figure}[tb]
    \centering
    \includegraphics[width = 3.3in]{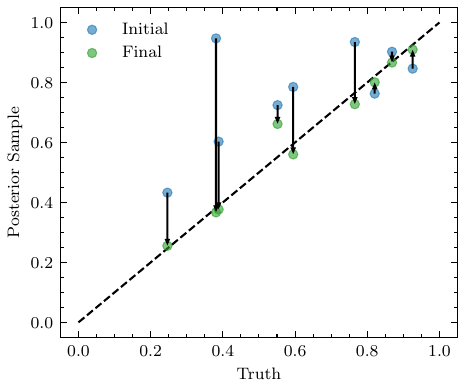}
    \caption{Progression of axis ratio, computed using AstroPhot ~\citep{Stone2023} with a Sersic light profile, for the posterior samples in \Fig{main_plot} as a function of the values computed for the ground truth galaxies. The x-axis represents the computed axis ratio for the ground truth samples, while the y-axis shows the axis ratios computed for the initial (blue) and final (green) posterior samples $\mathbf{x} \sim p_{\theta_\alpha}(\mathbf{x} \mid \mathbf{y})$. Each point compares the predicted axis ratio to the true value, with arrows indicating improvement between the initial and the final posterior samples. The dashed line represents the ideal scenario where the predicted ratios perfectly match the truth. This plot demonstrates a clear improvement from initial to final samples, with the final points consistently closer to the ideal line, indicating a significant reduction in bias and better alignment with the ground truth.}
    \label{fig:axisratio}
\end{figure}

A classic measure for galaxies is the axis ratio ($q$), defined as the ratio of the semi-minor over semi-major axis lengths. 
This gives a sense of the roundness of a galaxy ($q=1$ is circular, $q<1$ is elliptical) and is known to have a different observed distribution for elliptical galaxies and spirals.
As a concrete test of the bias introduced by using a misspecified prior, in \Fig{axisratio} we compare $q$ for the ground truth samples and the posterior samples shown in \Fig{main_plot}.
In all cases, the $q$ values converge towards the ground truth values for the updated posterior sampling.
Axis ratios are computed using AstroPhot assuming a Sersic light profile~\citep{Stone2023}, though the results are virtually unchanged for other light profile choices.

\subsubsection{Residuals Improvement}

\begin{figure*}[tb]
    \centering
    \includegraphics[width = 6.8in]{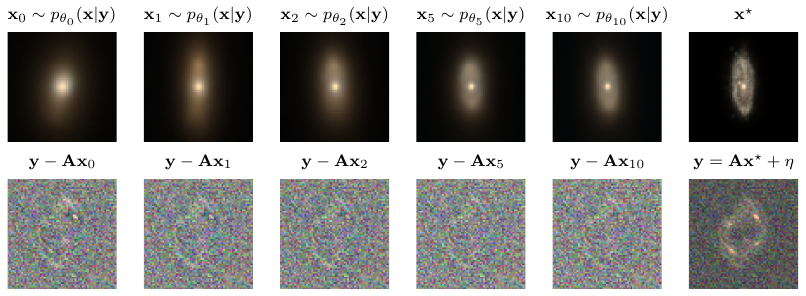}
    \caption{Posterior samples $\mathbf{x}_i \sim p_{\theta_i}(\mathbf{x} \mid \mathbf{y})$ and residuals $A\mathbf{x}_i - \mathbf{y}$, given the same observation $\mathbf{y}$, across updates in the galaxy experiment with observational noise of $\sigma_{\boldsymbol{\eta}} = 0.5$. The top row presents posterior samples $\mathbf{x}_i$ from their respective distributions, including the ground truth $\mathbf{x}^\star$. The bottom row shows the residuals $\mathbf{y} - A\mathbf{x}_i$ for each sample, with the last panel showing the mock observation $\mathbf{y} = A\mathbf{x}^\star + \boldsymbol{\eta}$ where $\boldsymbol{\eta}$ is the noise realization. The sequence illustrates that reconstructions increasingly approximate the ground truth over iterations, and the residuals evolve to resemble Gaussian noise, further detailed in \Fig{mean_llk} (bottom), which discusses the mean log-likelihood of the residuals.}
    \label{fig:single_posterior}
\end{figure*}

In \Fig{single_posterior}, we can observe the progress in reconstruction and residuals for a single observation using different priors $p_{\theta_i}(\mathbf{x})$. After varying numbers of iterations, the residuals increasingly resemble pure noise, and the reconstructions approach the ground truth $\mathbf{x}^\star$. Progress is more evident between the initial and the first updated prior, and less evident afterwards. This is also observed in the $\chi^2_{PQM}$ values and the mean likelihood of the residuals in \Fig{mean_llk}.

\subsubsection{TARP: Posterior Sampling Coverage Test}

Across all experiments, the posterior sampling procedure $\mathcal{F}$ is a Predictor-Corrector SDE solver, with a different number of steps. When doing posterior sampling, we use the convolved likelihood approximation. Since approximations are involved, and the problem is discretized with a certain number of steps, it is important to test the correctness of $\mathcal{F}$.

We choose to use TARP \citep{lemos2023-tarp}, a sample-based method to estimate coverage probabilities of generative posterior estimators. It has been shown that passing this test is a necessary and sufficient condition for the accuracy of $\mathcal{F}$. We perform several tests for experiments with both MNIST and galaxies. These tests can be conducted either using the test set of the dataset used to train the true distribution $p_{\theta^\star}$, or with samples from the true distribution since it is defined to be the target.

\begin{figure}[tb]
    \centering
    \includegraphics[width = 3in]{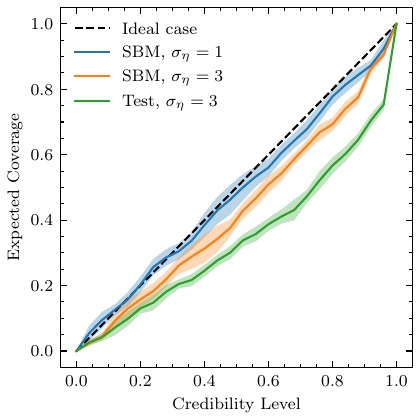}
    \caption{Coverage test for the posterior sampling procedure $\mathcal{F}$ using TARP~\citep{lemos2023-tarp} for the galaxy experiment. Credibility levels are plotted against expected coverage, comparing results from mock observations created with samples from the Score-Based Model (SBM) prior under different noise levels: $\sigma_{\boldsymbol{\eta}} = 1$ (blue) and $\sigma_{\boldsymbol{\eta}} = 3$ (orange), along with mock observations from the test set with noise $\sigma_{\boldsymbol{\eta}} = 3$ (green). The dashed line represents the ideal case where expected coverage matches credibility levels perfectly. Posterior sampling is exact when we have the correct prior (here enforced by creating mock observations with samples from the prior), and with a specific noise level. Results vary with different noise levels or a misspecified prior, indicating $\mathcal{F}$'s sensitivity to these factors.}
    \label{fig:tarp}
\end{figure}

For the experiment with galaxies, when using a Predictor-Corrector solver with $1\,024$ steps (one corrector step per predictor step) as $\mathcal{F}$, posterior sampling is exact, as shown in \Fig{tarp}. This uses the correct prior (SBM) and observational noise of $\sigma_{\boldsymbol{\eta}} = 1$. However, increasing the level of observational noise to $\sigma_{\boldsymbol{\eta}} = 3$ makes the procedure biased, indicating the limit of the approximations used for $\mathcal{F}$. Furthermore, when using the test set instead of samples from the prior, again with $\sigma_{\boldsymbol{\eta}} = 3$, we obtain more biased results. This could be because the SBM is not in distribution with the test set, possibly due to model capacity or imperfect learning.

When using $\mathcal{F}$ with a misspecified prior, the test shows an important bias, as it is sensitive to the correct prior. Nonetheless, we use $\mathcal{F}$ as is for the updates. To run TARP, we simulate $256$ observations and obtain $256$ posterior samples from each one.

\subsection{Limitations}

One of the main limitations of our current framework is that it requires a substantial amount of data and compute resources. The iterative retraining of SBM from scratch alone can be an important computational burden, as such models typically take days to train even on modern GPUs. 

Since the improvement per update slows down drastically after a few iterations, an argument can be made to generally keep the number of iterations low. However, \citet{vetter2024-sourcerer} have shown that even in the infinite information limit, a single update will not necessarily converge to the true prior distribution. The necessary number of steps $M$ before convergence depends on several factors, such as the characteristics of the initial and target distributions (e.g., the presence or absence of high-frequency features) and the differences between them, such as missing modes. For example, in \App{reverse_experiment}, we performed the same experiment with galaxy sources but in the reverse direction: the initial prior $p_{\theta_0}(\mathbf{x})$ was trained on the spiral galaxy dataset, while the true distribution $p_{\theta^\star}(\mathbf{x})$ was trained on the elliptical galaxy dataset. We observed faster convergence, requiring only $M = 6$ iterations. The initial prior $p_{\theta_0}(\mathbf{x})$ contains information about structures at more scales and only needs to \textit{forget} features, rather than having to \textit{learn} new ones. This is similar to what we observed in the MNIST experiment (\Sec{mnist_exp}), where digits are forgotten in a single update but require several iterations to be learned. We hypothesize that, in general, forgetting features is easier than learning new ones, though this would need to be quantified in future work. The noise level of the observations also affects the convergence speed, as well as how similar the final prior $p_{\theta_M}(\mathbf{x})$ is to the target distribution $p_{\theta^\star}(\mathbf{x})$ since the noise level affects the amount of information available in each sample. This is shown, for example, in \Fig{pqm_mnist} in \App{details_mnist}.

Furthermore, sampling from an SBM (either prior or posterior sampling) is computationally expensive. For example, producing $1\,024$ galaxy samples takes approximately 2 hours (wall-time) on an A100 GPU with 32 GB of VRAM allocated. Another important factor is the dimensionality of the data (MNIST being $\mathbf{x} \in \mathbb{R}^{28 \times 28}$ and galaxies $\mathbf{x} \in \mathbb{R}^{3 \times 64 \times 64}$), as it affects the complexity of the model $\mathbf{s}_\theta(\mathbf{x}, t)$ and the compute time across all steps of the procedure.

One possible improvement is to use fine-tuning techniques to leverage the initially trained SBM, significantly reducing the computational burden of each update. In future work, we plan to explore the use of LoRA weights \citep{LORA} for this purpose.

Finally, it is important to mention that in this work, we assumed that the physical forward model $A$, determined by the lensing configuration, is known. However, this is typically not the case with real data. Jointly sampling the lensing parameters and the pixelated source in the presence of real (potentially non-Gaussian) noise would be necessary to apply this approach to real data. Several works in the literature address this problem configuration, known as blind inversion, in the context of score-based models \citep[e.g.,][]{murata_gibbsddrm_2023, chung_parallel_2023}. We plan to explore these approaches in future work.

\section{Conclusion}

\label{sec:discussion}

In recent years, score-based generative models have been used extensively in inverse problem settings, both in computer science and astronomy. A particular problem of interest for astronomy is how to train an SBM when no data is available a priori to characterize prior distributions for inverse problems. A natural corollary to this question is how to update a prior that is severely misspecified for an inference task when data from other sources is used to encode the prior in those cases.

In this work, we have introduced an algorithm that addresses the misspecification of a prior SBM by updating it using only partial and noisy observations. This method allows us to accurately learn a distribution over high-dimensional spaces (images). We have demonstrated empirically that this method can learn new features and modes in linear inverse problem settings. We have also shown empirical metrics that suggest that our method converges to a stationary distribution with a marginal likelihood equal to that of the true, underlying data-generating process. 

Such a method is of high value considering the volume of partially corrupted observations currently available and upcoming in large surveys of the sky like the \textit{Euclid} space telescope \citep{Euclid} and the Vera C. Rubin Observatory \citep{LSST}. 
Extracting information from these surveys and encoding it in SBM neural network for future inference tasks is an important subject for the development of computational imaging techniques in astronomy. In future works, we thus plan to integrate real observations into our framework to demonstrate the usage of our method in realistic settings.

\section*{Acknowledgments}
This work is partially supported by Schmidt Futures, a philanthropic initiative founded by Eric and Wendy Schmidt as part of the Virtual Institute for Astrophysics (VIA). The work is in part supported by computational resources provided by Calcul Quebec and the Digital Research Alliance of Canada. Y.H. and L.P.-L. acknowledge support from the Canada Research Chairs Program, the National Sciences and Engineering Council of Canada through grants RGPIN-2020-05073 and 05102, and the Fonds de recherche du Québec through grants CFQCU-2024-348060, 2022-NC-301305 and 300397. C.S. acknowledges the support of an NSERC Postdoctoral Fellowship and a CITA National Fellowship.

\newpage
\bibliographystyle{aasjournal}
\bibliography{bib}

\begin{thebibliography}{}
\expandafter\ifx\csname natexlab\endcsname\relax\def\natexlab#1{#1}\fi
\providecommand{\url}[1]{\href{#1}{#1}}
\providecommand{\dodoi}[1]{doi:~\href{http://doi.org/#1}{\nolinkurl{#1}}}
\providecommand{\doeprint}[1]{\href{http://ascl.net/#1}{\nolinkurl{http://ascl.net/#1}}}
\providecommand{\doarXiv}[1]{\href{https://arxiv.org/abs/#1}{\nolinkurl{https://arxiv.org/abs/#1}}}

\bibitem[{{Abdurro'uf} {et~al.}(2022){Abdurro'uf}, {Accetta}, {Aerts}, {Silva Aguirre}, {Ahumada}, {Ajgaonkar}, {Filiz Ak}, {Alam}, {Allende Prieto}, {Almeida}, {Anders}, {Anderson}, {Andrews}, {Anguiano}, {Aquino-Ort{\'\i}z}, {Arag{\'o}n-Salamanca}, {Argudo-Fern{\'a}ndez}, {Ata}, {Aubert}, {Avila-Reese}, {Badenes}, {Barb{\'a}}, {Barger}, {Barrera-Ballesteros}, {Beaton}, {Beers}, {Belfiore}, {Bender}, {Bernardi}, {Bershady}, {Beutler}, {Bidin}, {Bird}, {Bizyaev}, {Blanc}, {Blanton}, {Boardman}, {Bolton}, {Boquien}, {Borissova}, {Bovy}, {Brandt}, {Brown}, {Brownstein}, {Brusa}, {Buchner}, {Bundy}, {Burchett}, {Bureau}, {Burgasser}, {Cabang}, {Campbell}, {Cappellari}, {Carlberg}, {Wanderley}, {Carrera}, {Cash}, {Chen}, {Chen}, {Cherinka}, {Chiappini}, {Choi}, {Chojnowski}, {Chung}, {Clerc}, {Cohen}, {Comerford}, {Comparat}, {da Costa}, {Covey}, {Crane}, {Cruz-Gonzalez}, {Culhane}, {Cunha}, {Dai}, {Damke}, {Darling}, {Davidson}, {Davies}, {Dawson}, {De Lee}, {Diamond-Stanic}, {Cano-D{\'\i}az}, {S{\'a}nchez},
  {Donor}, {Duckworth}, {Dwelly}, {Eisenstein}, {Elsworth}, {Emsellem}, {Eracleous}, {Escoffier}, {Fan}, {Farr}, {Feng}, {Fern{\'a}ndez-Trincado}, {Feuillet}, {Filipp}, {Fillingham}, {Frinchaboy}, {Fromenteau}, {Galbany}, {Garc{\'\i}a}, {Garc{\'\i}a-Hern{\'a}ndez}, {Ge}, {Geisler}, {Gelfand}, {G{\'e}ron}, {Gibson}, {Goddy}, {Godoy-Rivera}, {Grabowski}, {Green}, {Greener}, {Grier}, {Griffith}, {Guo}, {Guy}, {Hadjara}, {Harding}, {Hasselquist}, {Hayes}, {Hearty}, {Hern{\'a}ndez}, {Hill}, {Hogg}, {Holtzman}, {Horta}, {Hsieh}, {Hsu}, {Hsu}, {Huber}, {Huertas-Company}, {Hutchinson}, {Hwang}, {Ibarra-Medel}, {Chitham}, {Ilha}, {Imig}, {Jaekle}, {Jayasinghe}, {Ji}, {Johnson}, {Jones}, {J{\"o}nsson}, {Katkov}, {Khalatyan}, {Kinemuchi}, {Kisku}, {Knapen}, {Kneib}, {Kollmeier}, {Kong}, {Kounkel}, {Kreckel}, {Krishnarao}, {Lacerna}, {Lane}, {Langgin}, {Lavender}, {Law}, {Lazarz}, {Leung}, {Leung}, {Lewis}, {Li}, {Li}, {Lian}, {Liang}, {Lin}, {Lin}, {Lin}, {Lintott}, {Long}, {Longa-Pe{\~n}a}, {L{\'o}pez-Cob{\'a}}, {Lu},
  {Lundgren}, {Luo}, {Mackereth}, {de la Macorra}, {Mahadevan}, {Majewski}, {Manchado}, {Mandeville}, {Maraston}, {Margalef-Bentabol}, {Masseron}, {Masters}, {Mathur}, {McDermid}, {Mckay}, {Merloni}, {Merrifield}, {Meszaros}, {Miglio}, {Di Mille}, {Minniti}, {Minsley}, {Monachesi}, {Moon}, {Mosser}, {Mulchaey}, {Muna}, {Mu{\~n}oz}, {Myers}, {Myers}, {Nadathur}, {Nair}, {Nandra}, {Neumann}, {Newman}, {Nidever}, {Nikakhtar}, {Nitschelm}, {O'Connell}, {Garma-Oehmichen}, {Luan Souza de Oliveira}, {Olney}, {Oravetz}, {Ortigoza-Urdaneta}, {Osorio}, {Otter}, {Pace}, {Padilla}, {Pan}, {Pan}, {Parikh}, {Parker}, {Peirani}, {Pe{\~n}a Ram{\'\i}rez}, {Penny}, {Percival}, {Perez-Fournon}, {Pinsonneault}, {Poidevin}, {Poovelil}, {Price-Whelan}, {B{\'a}rbara de Andrade Queiroz}, {Raddick}, {Ray}, {Rembold}, {Riddle}, {Riffel}, {Riffel}, {Rix}, {Robin}, {Rodr{\'\i}guez-Puebla}, {Roman-Lopes}, {Rom{\'a}n-Z{\'u}{\~n}iga}, {Rose}, {Ross}, {Rossi}, {Rubin}, {Salvato}, {S{\'a}nchez}, {S{\'a}nchez-Gallego}, {Sanderson}, {Santana
  Rojas}, {Sarceno}, {Sarmiento}, {Sayres}, {Sazonova}, {Schaefer}, {Schiavon}, {Schlegel}, {Schneider}, {Schultheis}, {Schwope}, {Serenelli}, {Serna}, {Shao}, {Shapiro}, {Sharma}, {Shen}, {Shetrone}, {Shu}, {Simon}, {Skrutskie}, {Smethurst}, {Smith}, {Sobeck}, {Spoo}, {Sprague}, {Stark}, {Stassun}, {Steinmetz}, {Stello}, {Stone-Martinez}, {Storchi-Bergmann}, {Stringfellow}, {Stutz}, {Su}, {Taghizadeh-Popp}, {Talbot}, {Tayar}, {Telles}, {Teske}, {Thakar}, {Theissen}, {Tkachenko}, {Thomas}, {Tojeiro}, {Hernandez Toledo}, {Troup}, {Trump}, {Trussler}, {Turner}, {Tuttle}, {Unda-Sanzana}, {V{\'a}zquez-Mata}, {Valentini}, {Valenzuela}, {Vargas-Gonz{\'a}lez}, {Vargas-Maga{\~n}a}, {Alfaro}, {Villanova}, {Vincenzo}, {Wake}, {Warfield}, {Washington}, {Weaver}, {Weijmans}, {Weinberg}, {Weiss}, {Westfall}, {Wild}, {Wilde}, {Wilson}, {Wilson}, {Wilson}, {Wolf}, {Wood-Vasey}, {Yan}, {Zamora}, {Zasowski}, {Zhang}, {Zhao}, {Zheng}, {Zheng}, \& {Zhu}}]{SDSS-DR17}
{Abdurro'uf}, {Accetta}, K., {Aerts}, C., {et~al.} 2022, The Astrophysical Journal Supplement Series, 259, 35, \dodoi{10.3847/1538-4365/ac4414}

\bibitem[{{Adam} {et~al.}(2022){Adam}, {Coogan}, {Malkin}, {Legin}, {Perreault-Levasseur}, {Hezaveh}, \& {Bengio}}]{Adam2022}
{Adam}, A., {Coogan}, A., {Malkin}, N., {et~al.} 2022, in Machine Learning and the Physical Sciences Workshop, NeurIPS 2022.
\newblock \doarXiv{2211.03812}

\bibitem[{Adam {et~al.}(2023)Adam, Stone, Bottrell, Legin, Hezaveh, \& Perreault-Levasseur}]{Adam2023}
Adam, A., Stone, C., Bottrell, C., {et~al.} 2023, in Machine Learning and the Physical Sciences Workshop, NeurIPS 2023.
\newblock \doarXiv{2311.18002}

\bibitem[{{Aihara} {et~al.}(2018){Aihara}, {Arimoto}, {Armstrong}, {Arnouts}, {Bahcall}, {Bickerton}, {Bosch}, {Bundy}, {Capak}, {Chan}, {Chiba}, {Coupon}, {Egami}, {Enoki}, {Finet}, {Fujimori}, {Fujimoto}, {Furusawa}, {Furusawa}, {Goto}, {Goulding}, {Greco}, {Greene}, {Gunn}, {Hamana}, {Harikane}, {Hashimoto}, {Hattori}, {Hayashi}, {Hayashi}, {He{\l}miniak}, {Higuchi}, {Hikage}, {Ho}, {Hsieh}, {Huang}, {Huang}, {Ikeda}, {Imanishi}, {Inoue}, {Iwasawa}, {Iwata}, {Jaelani}, {Jian}, {Kamata}, {Karoji}, {Kashikawa}, {Katayama}, {Kawanomoto}, {Kayo}, {Koda}, {Koike}, {Kojima}, {Komiyama}, {Konno}, {Koshida}, {Koyama}, {Kusakabe}, {Leauthaud}, {Lee}, {Lin}, {Lin}, {Lupton}, {Mandelbaum}, {Matsuoka}, {Medezinski}, {Mineo}, {Miyama}, {Miyatake}, {Miyazaki}, {Momose}, {More}, {More}, {Moritani}, {Moriya}, {Morokuma}, {Mukae}, {Murata}, {Murayama}, {Nagao}, {Nakata}, {Niida}, {Niikura}, {Nishizawa}, {Obuchi}, {Oguri}, {Oishi}, {Okabe}, {Okamoto}, {Okura}, {Ono}, {Onodera}, {Onoue}, {Osato}, {Ouchi}, {Price}, {Pyo},
  {Sako}, {Sawicki}, {Shibuya}, {Shimasaku}, {Shimono}, {Shirasaki}, {Silverman}, {Simet}, {Speagle}, {Spergel}, {Strauss}, {Sugahara}, {Sugiyama}, {Suto}, {Suyu}, {Suzuki}, {Tait}, {Takada}, {Takata}, {Tamura}, {Tanaka}, {Tanaka}, {Tanaka}, {Tanaka}, {Terai}, {Terashima}, {Toba}, {Tominaga}, {Toshikawa}, {Turner}, {Uchida}, {Uchiyama}, {Umetsu}, {Uraguchi}, {Urata}, {Usuda}, {Utsumi}, {Wang}, {Wang}, {Wong}, {Yabe}, {Yamada}, {Yamanoi}, {Yasuda}, {Yeh}, {Yonehara}, \& {Yuma}}]{HSCSurvey}
{Aihara}, H., {Arimoto}, N., {Armstrong}, R., {et~al.} 2018, PASJ, 70, S4, \dodoi{10.1093/pasj/psx066}

\bibitem[{Anderson(1982)}]{Anderson1982}
Anderson, B.~D. 1982, Stochastic Processes and their Applications, 12, 313, \dodoi{https://doi.org/10.1016/0304-4149(82)90051-5}

\bibitem[{Bertrand {et~al.}(2024)Bertrand, Bose, Duplessis, Jiralerspong, \& Gidel}]{bertrand2024on}
Bertrand, Q., Bose, J., Duplessis, A., Jiralerspong, M., \& Gidel, G. 2024, in The Twelfth International Conference on Learning Representations.
\newblock \url{https://openreview.net/forum?id=JORAfH2xFd}

\bibitem[{{Blanton} {et~al.}(2017){Blanton}, {Bershady}, {Abolfathi}, {Albareti}, {Allende Prieto}, {Almeida}, {Alonso-Garc{\'\i}a}, {Anders}, {Anderson}, {Andrews}, {Aquino-Ort{\'\i}z}, {Arag{\'o}n-Salamanca}, {Argudo-Fern{\'a}ndez}, {Armengaud}, {Aubourg}, {Avila-Reese}, {Badenes}, {Bailey}, {Barger}, {Barrera-Ballesteros}, {Bartosz}, {Bates}, {Baumgarten}, {Bautista}, {Beaton}, {Beers}, {Belfiore}, {Bender}, {Berlind}, {Bernardi}, {Beutler}, {Bird}, {Bizyaev}, {Blanc}, {Blomqvist}, {Bolton}, {Boquien}, {Borissova}, {van den Bosch}, {Bovy}, {Brandt}, {Brinkmann}, {Brownstein}, {Bundy}, {Burgasser}, {Burtin}, {Busca}, {Cappellari}, {Delgado Carigi}, {Carlberg}, {Carnero Rosell}, {Carrera}, {Chanover}, {Cherinka}, {Cheung}, {G{\'o}mez Maqueo Chew}, {Chiappini}, {Choi}, {Chojnowski}, {Chuang}, {Chung}, {Cirolini}, {Clerc}, {Cohen}, {Comparat}, {da Costa}, {Cousinou}, {Covey}, {Crane}, {Croft}, {Cruz-Gonzalez}, {Garrido Cuadra}, {Cunha}, {Damke}, {Darling}, {Davies}, {Dawson}, {de la Macorra}, {Dell'Agli}, {De
  Lee}, {Delubac}, {Di Mille}, {Diamond-Stanic}, {Cano-D{\'\i}az}, {Donor}, {Downes}, {Drory}, {du Mas des Bourboux}, {Duckworth}, {Dwelly}, {Dyer}, {Ebelke}, {Eigenbrot}, {Eisenstein}, {Emsellem}, {Eracleous}, {Escoffier}, {Evans}, {Fan}, {Fern{\'a}ndez-Alvar}, {Fernandez-Trincado}, {Feuillet}, {Finoguenov}, {Fleming}, {Font-Ribera}, {Fredrickson}, {Freischlad}, {Frinchaboy}, {Fuentes}, {Galbany}, {Garcia-Dias}, {Garc{\'\i}a-Hern{\'a}ndez}, {Gaulme}, {Geisler}, {Gelfand}, {Gil-Mar{\'\i}n}, {Gillespie}, {Goddard}, {Gonzalez-Perez}, {Grabowski}, {Green}, {Grier}, {Gunn}, {Guo}, {Guy}, {Hagen}, {Hahn}, {Hall}, {Harding}, {Hasselquist}, {Hawley}, {Hearty}, {Gonzalez Hern{\'a}ndez}, {Ho}, {Hogg}, {Holley-Bockelmann}, {Holtzman}, {Holzer}, {Huehnerhoff}, {Hutchinson}, {Hwang}, {Ibarra-Medel}, {da Silva Ilha}, {Ivans}, {Ivory}, {Jackson}, {Jensen}, {Johnson}, {Jones}, {J{\"o}nsson}, {Jullo}, {Kamble}, {Kinemuchi}, {Kirkby}, {Kitaura}, {Klaene}, {Knapp}, {Kneib}, {Kollmeier}, {Lacerna}, {Lane}, {Lang}, {Law},
  {Lazarz}, {Lee}, {Le Goff}, {Liang}, {Li}, {Li}, {Lian}, {Lima}, {Lin}, {Lin}, {Bertran de Lis}, {Liu}, {de Icaza Lizaola}, {Long}, {Lucatello}, {Lundgren}, {MacDonald}, {Deconto Machado}, {MacLeod}, {Mahadevan}, {Geimba Maia}, {Maiolino}, {Majewski}, {Malanushenko}, {Malanushenko}, {Manchado}, {Mao}, {Maraston}, {Marques-Chaves}, {Masseron}, {Masters}, {McBride}, {McDermid}, {McGrath}, {McGreer}, {Medina Pe{\~n}a}, {Melendez}, {Merloni}, {Merrifield}, {Meszaros}, {Meza}, {Minchev}, {Minniti}, {Miyaji}, {More}, {Mulchaey}, {M{\"u}ller-S{\'a}nchez}, {Muna}, {Munoz}, {Myers}, {Nair}, {Nandra}, {Correa do Nascimento}, {Negrete}, {Ness}, {Newman}, {Nichol}, {Nidever}, {Nitschelm}, {Ntelis}, {O'Connell}, {Oelkers}, {Oravetz}, {Oravetz}, {Pace}, {Padilla}, {Palanque-Delabrouille}, {Alonso Palicio}, {Pan}, {Parejko}, {Parikh}, {P{\^a}ris}, {Park}, {Patten}, {Peirani}, {Pellejero-Ibanez}, {Penny}, {Percival}, {Perez-Fournon}, {Petitjean}, {Pieri}, {Pinsonneault}, {Pisani}, {Poleski}, {Prada}, {Prakash}, {Queiroz},
  {Raddick}, {Raichoor}, {Barboza Rembold}, {Richstein}, {Riffel}, {Riffel}, {Rix}, {Robin}, {Rockosi}, {Rodr{\'\i}guez-Torres}, {Roman-Lopes}, {Rom{\'a}n-Z{\'u}{\~n}iga}, {Rosado}, {Ross}, {Rossi}, {Ruan}, {Ruggeri}, {Rykoff}, {Salazar-Albornoz}, {Salvato}, {S{\'a}nchez}, {Aguado}, {S{\'a}nchez-Gallego}, {Santana}, {Santiago}, {Sayres}, {Schiavon}, {da Silva Schimoia}, {Schlafly}, {Schlegel}, {Schneider}, {Schultheis}, {Schuster}, {Schwope}, {Seo}, {Shao}, {Shen}, {Shetrone}, {Shull}, {Simon}, {Skinner}, {Skrutskie}, {Slosar}, {Smith}, {Sobeck}, {Sobreira}, {Somers}, {Souto}, {Stark}, {Stassun}, {Stauffer}, {Steinmetz}, {Storchi-Bergmann}, {Streblyanska}, {Stringfellow}, {Su{\'a}rez}, {Sun}, {Suzuki}, {Szigeti}, {Taghizadeh-Popp}, {Tang}, {Tao}, {Tayar}, {Tembe}, {Teske}, {Thakar}, {Thomas}, {Thompson}, {Tinker}, {Tissera}, {Tojeiro}, {Hernandez Toledo}, {de la Torre}, {Tremonti}, {Troup}, {Valenzuela}, {Martinez Valpuesta}, {Vargas-Gonz{\'a}lez}, {Vargas-Maga{\~n}a}, {Vazquez}, {Villanova}, {Vivek}, {Vogt},
  {Wake}, {Walterbos}, {Wang}, {Weaver}, {Weijmans}, {Weinberg}, {Westfall}, {Whelan}, {Wild}, {Wilson}, {Wood-Vasey}, {Wylezalek}, {Xiao}, {Yan}, {Yang}, {Ybarra}, {Y{\`e}che}, {Zakamska}, {Zamora}, {Zarrouk}, {Zasowski}, {Zhang}, {Zhao}, {Zheng}, {Zheng}, {Zhou}, {Zhou}, {Zhu}, {Zoccali}, \& {Zou}}]{Blanton2017-SDSS-IV}
{Blanton}, M.~R., {Bershady}, M.~A., {Abolfathi}, B., {et~al.} 2017, The Astronomical Journal, 154, 28, \dodoi{10.3847/1538-3881/aa7567}

\bibitem[{{Blum} {et~al.}(2022){Blum}, {Digel}, {Drlica-Wagner}, {Habib}, {Heitmann}, {Ishak}, {Jha}, {Kahn}, {Mandelbaum}, {Marshall}, {Newman}, {Roodman}, \& {Stubbs}}]{LSST}
{Blum}, B., {Digel}, S.~W., {Drlica-Wagner}, A., {et~al.} 2022, arXiv e-prints, arXiv:2203.07220, \dodoi{10.48550/arXiv.2203.07220}

\bibitem[{{Bottrell} {et~al.}(2024){Bottrell}, {Yesuf}, {Popping}, {Omori}, {Tang}, {Ding}, {Pillepich}, {Nelson}, {Eisert}, {Gao}, {Goulding}, {Kalita}, {Luo}, {Greene}, {Shi}, \& {Silverman}}]{Bottrell2024}
{Bottrell}, C., {Yesuf}, H.~M., {Popping}, G., {et~al.} 2024, MNRAS, 527, 6506, \dodoi{10.1093/mnras/stad2971}

\bibitem[{Camps \& Baes(2020)}]{Camps2020}
Camps, P., \& Baes, M. 2020, Astronomy and Computing, 31, 100381, \dodoi{https://doi.org/10.1016/j.ascom.2020.100381}

\bibitem[{Casella(1985)}]{introEmpiricalBayes}
Casella, G. 1985, The American Statistician, 39, 83, \dodoi{10.1080/00031305.1985.10479400}

\bibitem[{Chung {et~al.}(2023{\natexlab{a}})Chung, Kim, Kim, \& Ye}]{chung_parallel_2023}
Chung, H., Kim, J., Kim, S., \& Ye, J.~C. 2023{\natexlab{a}}, in 2023 {IEEE}/{CVF} {Conference} on {Computer} {Vision} and {Pattern} {Recognition} ({CVPR}) (Vancouver, BC, Canada: IEEE), 6059--6069, \dodoi{10.1109/CVPR52729.2023.00587}

\bibitem[{Chung {et~al.}(2023{\natexlab{b}})Chung, Kim, Mccann, Klasky, \& Ye}]{Chung2023inverse}
Chung, H., Kim, J., Mccann, M.~T., Klasky, M.~L., \& Ye, J.~C. 2023{\natexlab{b}}, in The Eleventh International Conference on Learning Representations.
\newblock \url{https://openreview.net/forum?id=OnD9zGAGT0k}

\bibitem[{Chung \& Ye(2022)}]{Chung2022mri}
Chung, H., \& Ye, J.~C. 2022, Medical Image Analysis, 80, 102479, \dodoi{https://doi.org/10.1016/j.media.2022.102479}

\bibitem[{{Collett}(2015)}]{Collett2015}
{Collett}, T.~E. 2015, \apj, 811, 20, \dodoi{10.1088/0004-637X/811/1/20}

\bibitem[{Consonni {et~al.}(2018)Consonni, Fouskakis, Liseo, \& Ntzoufras}]{consonni2018prior}
Consonni, G., Fouskakis, D., Liseo, B., \& Ntzoufras, I. 2018, Bayesian Analysis, 13, 627

\bibitem[{Dempster {et~al.}(1977)Dempster, Laird, \& Rubin}]{Dempster1977}
Dempster, A.~P., Laird, N.~M., \& Rubin, D.~B. 1977, Journal of the Royal Statistical Society: Series B (Methodological), 39, 1, \dodoi{10.1111/j.2517-6161.1977.tb01600.x}

\bibitem[{Dey {et~al.}(2019)Dey, Schlegel, Lang, Blum, Burleigh, Fan, Findlay, Finkbeiner, Herrera, Juneau, Landriau, Levi, McGreer, Meisner, Myers, Moustakas, Nugent, Patej, Schlafly, Walker, Valdes, Weaver, Yèche, Zou, Zhou, Abareshi, Abbott, Abolfathi, Aguilera, Alam, Allen, Alvarez, Annis, Ansarinejad, Aubert, Beechert, Bell, BenZvi, Beutler, Bielby, Bolton, Briceño, Buckley-Geer, Butler, Calamida, Carlberg, Carter, Casas, Castander, Choi, Comparat, Cukanovaite, Delubac, DeVries, Dey, Dhungana, Dickinson, Ding, Donaldson, Duan, Duckworth, Eftekharzadeh, Eisenstein, Etourneau, Fagrelius, Farihi, Fitzpatrick, Font-Ribera, Fulmer, Gänsicke, Gaztanaga, George, Gerdes, Gontcho, Gorgoni, Green, Guy, Harmer, Hernandez, Honscheid, Huang, James, Jannuzi, Jiang, Joyce, Karcher, Karkar, Kehoe, Jean-Paul, Kueter-Young, Lan, Lauer, Guillou, Suu, Lee, Lesser, Levasseur, Li, Mann, Marshall, Martínez-Vázquez, Martini, du~Mas~des Bourboux, McManus, Meier, Ménard, Metcalfe, Muñoz-Gutiérrez, Najita, Napier, Narayan,
  Newman, Nie, Nord, Norman, Olsen, Paat, Palanque-Delabrouille, Peng, Poppett, Poremba, Prakash, Rabinowitz, Raichoor, Rezaie, Robertson, Roe, Ross, Ross, Rudnick, Safonova, Saha, Sánchez, Savary, Schweiker, Scott, Seo, Shan, Silva, Slepian, Soto, Sprayberry, Staten, Stillman, Stupak, Summers, Tie, Tirado, Vargas-Magaña, Vivas, Wechsler, Williams, Yang, Yang, Yapici, Zaritsky, Zenteno, Zhang, Zhang, Zhou, \& Zhou}]{Dey_2019}
Dey, A., Schlegel, D.~J., Lang, D., {et~al.} 2019, The Astronomical Journal, 157, 168, \dodoi{10.3847/1538-3881/ab089d}

\bibitem[{Dhariwal \& Nichol(2021)}]{Dhariwal2021beatGAN}
Dhariwal, P., \& Nichol, A. 2021, in Advances in Neural Information Processing Systems, ed. M.~Ranzato, A.~Beygelzimer, Y.~Dauphin, P.~Liang, \& J.~W. Vaughan, Vol.~34 (Curran Associates, Inc.), 8780--8794.
\newblock \url{https://proceedings.neurips.cc/paper_files/paper/2021/file/49ad23d1ec9fa4bd8d77d02681df5cfa-Paper.pdf}

\bibitem[{Dia {et~al.}(2023)Dia, Yantovski-Barth, Adam, Bowles, Lemos, Scaife, Hezaveh, \& Perreault-Levasseur}]{Dia2023}
Dia, N., Yantovski-Barth, M.~J., Adam, A., {et~al.} 2023, in Machine Learning and the Physical Sciences Workshop, NeurIPS 2023.
\newblock \doarXiv{2311.18012}

\bibitem[{{Dinh} {et~al.}(2014){Dinh}, {Krueger}, \& {Bengio}}]{Dinh2014}
{Dinh}, L., {Krueger}, D., \& {Bengio}, Y. 2014, in Workshop, The Sixth International Conference on Learning Representations

\bibitem[{Dou \& Song(2024)}]{Dou2024inverse}
Dou, Z., \& Song, Y. 2024, in The Twelfth International Conference on Learning Representations.
\newblock \url{https://openreview.net/forum?id=tplXNcHZs1}

\bibitem[{{Drozdova} {et~al.}(2024){Drozdova}, {Kinakh}, {Bait}, {Taran}, {Lastufka}, {Dessauges-Zavadsky}, {Holotyak}, {Schaerer}, \& {Voloshynovskiy}}]{Drozdova2024}
{Drozdova}, M., {Kinakh}, V., {Bait}, O., {et~al.} 2024, A\&A, 683, A105, \dodoi{10.1051/0004-6361/202347948}

\bibitem[{{EHT Collaboration} {et~al.}(2019){EHT Collaboration}, {Akiyama}, {Alberdi}, {Alef}, {Asada}, {Azulay}, {Baczko}, {Ball}, {Balokovi{\'c}}, {Barrett}, {Bintley}, {Blackburn}, {Boland}, {Bouman}, {Bower}, {Bremer}, {Brinkerink}, {Brissenden}, {Britzen}, {Broderick}, {Broguiere}, {Bronzwaer}, {Byun}, {Carlstrom}, {Chael}, {Chan}, {Chatterjee}, {Chatterjee}, {Chen}, {Chen}, {Cho}, {Christian}, {Conway}, {Cordes}, {Crew}, {Cui}, {Davelaar}, {De Laurentis}, {Deane}, {Dempsey}, {Desvignes}, {Dexter}, {Doeleman}, {Eatough}, {Falcke}, {Fish}, {Fomalont}, {Fraga-Encinas}, {Friberg}, {Fromm}, {G{\'o}mez}, {Galison}, {Gammie}, {Garc{\'\i}a}, {Gentaz}, {Georgiev}, {Goddi}, {Gold}, {Gu}, {Gurwell}, {Hada}, {Hecht}, {Hesper}, {Ho}, {Ho}, {Honma}, {Huang}, {Huang}, {Hughes}, {Ikeda}, {Inoue}, {Issaoun}, {James}, {Jannuzi}, {Janssen}, {Jeter}, {Jiang}, {Johnson}, {Jorstad}, {Jung}, {Karami}, {Karuppusamy}, {Kawashima}, {Keating}, {Kettenis}, {Kim}, {Kim}, {Kim}, {Kino}, {Koay}, {Koch}, {Koyama}, {Kramer}, {Kramer},
  {Krichbaum}, {Kuo}, {Lauer}, {Lee}, {Li}, {Li}, {Lindqvist}, {Liu}, {Liuzzo}, {Lo}, {Lobanov}, {Loinard}, {Lonsdale}, {Lu}, {MacDonald}, {Mao}, {Markoff}, {Marrone}, {Marscher}, {Mart{\'\i}-Vidal}, {Matsushita}, {Matthews}, {Medeiros}, {Menten}, {Mizuno}, {Mizuno}, {Moran}, {Moriyama}, {Moscibrodzka}, {M{\"u}ller}, {Nagai}, {Nagar}, {Nakamura}, {Narayan}, {Narayanan}, {Natarajan}, {Neri}, {Ni}, {Noutsos}, {Okino}, {Olivares}, {Ortiz-Le{\'o}n}, {Oyama}, {{\"O}zel}, {Palumbo}, {Patel}, {Pen}, {Pesce}, {Pi{\'e}tu}, {Plambeck}, {PopStefanija}, {Porth}, {Prather}, {Preciado-L{\'o}pez}, {Psaltis}, {Pu}, {Ramakrishnan}, {Rao}, {Rawlings}, {Raymond}, {Rezzolla}, {Ripperda}, {Roelofs}, {Rogers}, {Ros}, {Rose}, {Roshanineshat}, {Rottmann}, {Roy}, {Ruszczyk}, {Ryan}, {Rygl}, {S{\'a}nchez}, {S{\'a}nchez-Arguelles}, {Sasada}, {Savolainen}, {Schloerb}, {Schuster}, {Shao}, {Shen}, {Small}, {Sohn}, {SooHoo}, {Tazaki}, {Tiede}, {Tilanus}, {Titus}, {Toma}, {Torne}, {Trent}, {Trippe}, {Tsuda}, {van Bemmel}, {van Langevelde},
  {van Rossum}, {Wagner}, {Wardle}, {Weintroub}, {Wex}, {Wharton}, {Wielgus}, {Wong}, {Wu}, {Young}, {Young}, {Younsi}, {Yuan}, {Yuan}, {Zensus}, {Zhao}, {Zhao}, {Zhu}, {Algaba}, {Allardi}, {Amestica}, {Bach}, {Beaudoin}, {Benson}, {Berthold}, {Blanchard}, {Blundell}, {Bustamente}, {Cappallo}, {Castillo-Dom{\'\i}nguez}, {Chang}, {Chang}, {Chang}, {Chen}, {Chilson}, {Chuter}, {C{\'o}rdova Rosado}, {Coulson}, {Crawford}, {Crowley}, {David}, {Derome}, {Dexter}, {Dornbusch}, {Dudevoir}, {Dzib}, {Eckert}, {Erickson}, {Everett}, {Faber}, {Farah}, {Fath}, {Folkers}, {Forbes}, {Freund}, {G{\'o}mez-Ruiz}, {Gale}, {Gao}, {Geertsema}, {Graham}, {Greer}, {Grosslein}, {Gueth}, {Halverson}, {Han}, {Han}, {Hao}, {Hasegawa}, {Henning}, {Hern{\'a}ndez-G{\'o}mez}, {Herrero-Illana}, {Heyminck}, {Hirota}, {Hoge}, {Huang}, {Impellizzeri}, {Jiang}, {Kamble}, {Keisler}, {Kimura}, {Kono}, {Kubo}, {Kuroda}, {Lacasse}, {Laing}, {Leitch}, {Li}, {Lin}, {Liu}, {Liu}, {Lu}, {Marson}, {Martin-Cocher}, {Massingill}, {Matulonis}, {McColl},
  {McWhirter}, {Messias}, {Meyer-Zhao}, {Michalik}, {Monta{\~n}a}, {Montgomerie}, {Mora-Klein}, {Muders}, {Nadolski}, {Navarro}, {Nguyen}, {Nishioka}, {Norton}, {Nystrom}, {Ogawa}, {Oshiro}, {Oyama}, {Padin}, {Parsons}, {Paine}, {Pe{\~n}alver}, {Phillips}, {Poirier}, {Pradel}, {Primiani}, {Raffin}, {Rahlin}, {Reiland}, {Risacher}, {Ruiz}, {S{\'a}ez-Mada{\'\i}n}, {Sassella}, {Schellart}, {Shaw}, {Silva}, {Shiokawa}, {Smith}, {Snow}, {Souccar}, {Sousa}, {Sridharan}, {Srinivasan}, {Stahm}, {Stark}, {Story}, {Timmer}, {Vertatschitsch}, {Walther}, {Wei}, {Whitehorn}, {Whitney}, {Woody}, {Wouterloot}, {Wright}, {Yamaguchi}, {Yu}, {Zeballos}, \& {Ziurys}}]{EHT}
{EHT Collaboration}, {Akiyama}, K., {Alberdi}, A., {et~al.} 2019, The Astrophysical Journal Letters, 875, L2, \dodoi{10.3847/2041-8213/ab0c96}

\bibitem[{{Euclid Collaboration} {et~al.}(2024){Euclid Collaboration}, {Mellier}, {Abdurro'uf}, {Acevedo Barroso}, {Ach{\'u}carro}, {Adamek}, {Adam}, {Addison}, {Aghanim}, {Aguena}, {Ajani}, {Akrami}, {Al-Bahlawan}, {Alavi}, {Albuquerque}, {Alestas}, {Alguero}, {Allaoui}, {Allen}, {Allevato}, {Alonso-Tetilla}, {Altieri}, {Alvarez-Candal}, {Amara}, {Amendola}, {Amiaux}, {Andika}, {Andreon}, {Andrews}, {Angora}, {Angulo}, {Annibali}, {Anselmi}, {Anselmi}, {Arcari}, {Archidiacono}, {Aric{\`o}}, {Arnaud}, {Arnouts}, {Asgari}, {Asorey}, {Atayde}, {Atek}, {Atrio-Barandela}, {Aubert}, {Aubourg}, {Auphan}, {Auricchio}, {Aussel}, {Aussel}, {Avelino}, {Avgoustidis}, {Avila}, {Awan}, {Azzollini}, {Baccigalupi}, {Bachelet}, {Bacon}, {Baes}, {Bagley}, {Bahr-Kalus}, {Balaguera-Antolinez}, {Balbinot}, {Balcells}, {Baldi}, {Baldry}, {Balestra}, {Ballardini}, {Ballester}, {Balogh}, {Ba{\~n}ados}, {Barbier}, {Bardelli}, {Barreiro}, {Barriere}, {Barros}, {Barthelemy}, {Bartolo}, {Basset}, {Battaglia}, {Battisti}, {Baugh},
  {Baumont}, {Bazzanini}, {Beaulieu}, {Beckmann}, {Belikov}, {Bel}, {Bellagamba}, {Bella}, {Bellini}, {Benabed}, {Bender}, {Benevento}, {Bennett}, {Benson}, {Bergamini}, {Bermejo-Climent}, {Bernardeau}, {Bertacca}, {Berthe}, {Berthier}, {Bethermin}, {Beutler}, {Bevillon}, {Bhargava}, {Bhatawdekar}, {Bisigello}, {Biviano}, {Blake}, {Blanchard}, {Blazek}, {Blot}, {Bosco}, {Bodendorf}, {Boenke}, {B{\"o}hringer}, {Bolzonella}, {Bonchi}, {Bonici}, {Bonino}, {Bonino}, {Bonvin}, {Bon}, {Booth}, {Borgani}, {Borlaff}, {Borsato}, {Bosco}, {Bose}, {Botticella}, {Boucaud}, {Bouche}, {Boucher}, {Boutigny}, {Bouvard}, {Bouy}, {Bowler}, {Bozza}, {Bozzo}, {Branchini}, {Brau-Nogue}, {Brekke}, {Bremer}, {Brescia}, {Breton}, {Brinchmann}, {Brinckmann}, {Brockley-Blatt}, {Brodwin}, {Brouard}, {Brown}, {Bruton}, {Bucko}, {Buddelmeijer}, {Buenadicha}, {Buitrago}, {Burger}, {Burigana}, {Busillo}, {Busonero}, {Cabanac}, {Cabayol-Garcia}, {Cagliari}, {Caillat}, {Caillat}, {Calabrese}, {Calabro}, {Calderone}, {Calura}, {Camacho
  Quevedo}, {Camera}, {Campos}, {Canas-Herrera}, {Candini}, {Cantiello}, {Capobianco}, {Cappellaro}, {Cappelluti}, {Cappi}, {Caputi}, {Cara}, {Carbone}, {Cardone}, {Carella}, {Carlberg}, {Carle}, {Carminati}, {Caro}, {Carrasco}, {Carretero}, {Carrilho}, {Carron Duque}, {Carry}, {Carvalho}, {Carvalho}, {Casas}, {Casas}, {Casenove}, {Casey}, {Cassata}, {Castander}, {Castelao}, {Castellano}, {Castiblanco}, {Castignani}, {Castro}, {Cavet}, {Cavuoti}, {Chabaud}, {Chambers}, {Charles}, {Charlot}, {Chartab}, {Chary}, {Chaumeil}, {Cho}, {Chon}, {Ciancetta}, {Ciliegi}, {Cimatti}, {Cimino}, {Cioni}, {Claydon}, {Cleland}, {Cl{\'e}ment}, {Clements}, {Clerc}, {Clesse}, {Codis}, {Cogato}, {Colbert}, {Cole}, {Coles}, {Collett}, {Collins}, {Colodro-Conde}, {Colombo}, {Combes}, {Conforti}, {Congedo}, {Conseil}, {Conselice}, {Contarini}, {Contini}, {Conversi}, {Cooray}, {Copin}, {Corasaniti}, {Corcho-Caballero}, {Corcione}, {Cordes}, {Corpace}, {Correnti}, {Costanzi}, {Costille}, {Courbin}, {Courcoult Mifsud}, {Courtois},
  {Cousinou}, {Covone}, {Cowell}, {Cragg}, {Cresci}, {Cristiani}, {Crocce}, {Cropper}, {E Crouzet}, {Csizi}, {Cuby}, {Cucchetti}, {Cucciati}, {Cuillandre}, {Cunha}, {Cuozzo}, {Daddi}, {D'Addona}, {Dafonte}, {Dagoneau}, {Dalessandro}, {Dalton}, {D'Amico}, {Dannerbauer}, {Danto}, {Das}, {Da Silva}, {da Silva}, {Daste}, {Davies}, {Davini}, {de Boer}, {Decarli}, {De Caro}, {Degaudenzi}, {Degni}, {de Jong}, {de la Bella}, {de la Torre}, {Delhaise}, {Delley}, {Delucchi}, {De Lucia}, {Denniston}, {De Paolis}, {De Petris}, {Derosa}, {Desai}, {Desjacques}, {Despali}, {Desprez}, {De Vicente-Albendea}, {Deville}, {Dias}, {D{\'\i}az-S{\'a}nchez}, {Diaz}, {Di Domizio}, {Diego}, {Di Ferdinando}, {Di Giorgio}, {Dimauro}, {Dinis}, {Dolag}, {Dolding}, {Dole}, {Dom{\'\i}nguez S{\'a}nchez}, {Dor{\'e}}, {Dournac}, {Douspis}, {Dreihahn}, {Droge}, {Dryer}, {Dubath}, {Duc}, {Ducret}, {Duffy}, {Dufresne}, {Duncan}, {Dupac}, {Duret}, {Durrer}, {Durret}, {Dusini}, {Ealet}, {Eggemeier}, {Eisenhardt}, {Elbaz}, {Elkhashab}, {Ellien},
  {Endicott}, {Enia}, {Erben}, {Escartin Vigo}, {Escoffier}, {Escudero Sanz}, {Essert}, {Ettori}, {Ezziati}, {Fabbian}, {Fabricius}, {Fang}, {Farina}, {Farina}, {Farinelli}, {Farrens}, {Faustini}, {Feltre}, {Ferguson}, {Ferrando}, {Ferrari}, {Ferr{\'e}-Mateu}, {Ferreira}, {Ferreras}, {Ferrero}, {Ferriol}, {Ferruit}, {Filleul}, {Finelli}, {Finkelstein}, {Finoguenov}, {Fiorini}, {Flentge}, {Focardi}, {Fonseca}, {Fontana}, {Fontanot}, {Fornari}, {Fosalba}, {Fossati}, {Fotopoulou}, {Fouchez}, {Fourmanoit}, {Frailis}, {Fraix-Burnet}, {Franceschi}, {Franco}, {Franzetti}, {Freihoefer}, {Frittoli}, {Frugier}, {Frusciante}, {Fumagalli}, {Fumagalli}, {Fumana}, {Fu}, {Gabarra}, {Galeotta}, {Galluccio}, {Ganga}, {Gao}, {Garc{\'\i}a-Bellido}, {Garcia}, {Gardner}, {Garilli}, {Gaspar-Venancio}, {Gasparetto}, {Gautard}, {Gavazzi}, {Gaztanaga}, {Genolet}, {Genova Santos}, {Gentile}, {George}, {Ghaffari}, {Giacomini}, {Gianotti}, {Gibb}, {Gillard}, {Gillis}, {Ginolfi}, {Giocoli}, {Girardi}, {Giri}, {Goh}, {G{\'o}mez-Alvarez},
  {Gonzalez}, {Gonzalez}, {Gonzalez}, {Gouyou Beauchamps}, {Gozaliasl}, {Gracia-Carpio}, {Grandis}, {Granett}, {Granvik}, {Grazian}, {Gregorio}, {Grenet}, {Grillo}, {Grupp}, {Gruppioni}, {Gruppuso}, {Guerbuez}, {Guerrini}, {Guidi}, {Guillard}, {Gutierrez}, {Guttridge}, {Guzzo}, {Gwyn}, {Haapala}, {Haase}, {Haddow}, {Hailey}, {Hall}, {Hall}, {Hamaus}, {Haridasu}, {Harnois-D{\'e}raps}, {Harper}, {Hartley}, {Hasinger}, {Hassani}, {Hatch}, {Haugan}, {H{\"a}u{\ss}ler}, {Heavens}, {Heisenberg}, {Helmi}, {Helou}, {Hemmati}, {Henares}, {Herent}, {Hern{\'a}ndez-Monteagudo}, {Heuberger}, {Hewett}, {Heydenreich}, {Hildebrandt}, {Hirschmann}, {Hjorth}, {Hoar}, {Hoekstra}, {Holland}, {Holliman}, {Holmes}, {Hook}, {Horeau}, {Hormuth}, {Hornstrup}, {Hosseini}, {Hu}, {Hudelot}, {Hudson}, {Huertas-Company}, {Huff}, {Hughes}, {Humphrey}, {Hunt}, {Huynh}, {Ibata}, {Ichikawa}, {Iglesias-Groth}, {Ilbert}, {Ili{\'c}}, {Ingoglia}, {Iodice}, {Israel}, {Israelsson}, {Izzo}, {Jablonka}, {Jackson}, {Jacobson}, {Jafariyazani}, {Jahnke},
  {Jansen}, {Jarvis}, {Jasche}, {Jauzac}, {Jeffrey}, {Jhabvala}, {Jimenez-Teja}, {Jimenez Mu{\~n}oz}, {Joachimi}, {Johansson}, {Joudaki}, {Jullo}, {Kajava}, {Kang}, {Kannawadi}, {Kansal}, {Karagiannis}, {K{\"a}rcher}, {Kashlinsky}, {Kazandjian}, {Keck}, {Keih{\"a}nen}, {Kerins}, {Kermiche}, {Khalil}, {Kiessling}, {Kiiveri}, {Kilbinger}, {Kim}, {King}, {Kirkpatrick}, {Kitching}, {Kluge}, {Knabenhans}, {Knapen}, {Knebe}, {Kneib}, {Kohley}, {Koopmans}, {Koskinen}, {Koulouridis}, {Kou}, {Kov{\'a}cs}, {Kova\{{\v{c}}\}i{\'c}}, {Kowalczyk}, {Koyama}, {Kraljic}, {Krause}, {Kruk}, {Kubik}, {Kuchner}, {Kuijken}, {K{\"u}mmel}, {Kunz}, {Kurki-Suonio}, {Lacasa}, {Lacey}, {La Franca}, {Lagarde}, {Lahav}, {Laigle}, {La Marca}, {La Marle}, {Lamine}, {Lam}, {Lan{\c{c}}on}, {Landt}, {Langer}, {Lapi}, {Larcheveque}, {Larsen}, {Lattanzi}, {Laudisio}, {Laugier}, {Laureijs}, {Lavaux}, {Lawrenson}, {Lazanu}, {Lazeyras}, {Le Boulc'h}, {Le Brun}, {Le Brun}, {Leclercq}, {Lee}, {Le Graet}, {Legrand}, {Leirvik}, {Le Jeune}, {Lembo}, {Le
  Mignant}, {Lepinzan}, {Lepori}, {Lesci}, {Lesgourgues}, {Leuzzi}, {Levi}, {Liaudat}, {Libet}, {Liebing}, {Ligori}, {Lilje}, {Lin}, {Linde}, {Linder}, {Lindholm}, {Linke}, {Li}, {Liu}, {Lloro}, {Lobo}, {Lodieu}, {Lombardi}, {Lombriser}, {Lonare}, {Longo}, {L{\'o}pez-Caniego}, {Lopez Lopez}, {Alvarez}, {Loureiro}, {Loveday}, {Lusso}, {Macias-Perez}, {Maciaszek}, {Magliocchetti}, {Magnard}, {Magnier}, {Magro}, {Mahler}, {Mainetti}, {Maino}, {Maiorano}, {Maiorano}, {Malavasi}, {Mamon}, {Mancini}, {Mandelbaum}, {Manera}, {Manj{\'o}n-Garc{\'\i}a}, {Mannucci}, {Mansutti}, {Manteiga Outeiro}, {Maoli}, {Maraston}, {Marcin}, {Marcos-Arenal}, {Margalef-Bentabol}, {Marggraf}, {Marinucci}, {Marinucci}, {Markovic}, {Marleau}, {Marpaud}, {Martignac}, {Mart{\'\i}n-Fleitas}, {Martin-Moruno}, {Martin}, {Martinelli}, {Martinet}, {Martin}, {Martins}, {Marulli}, {Massari}, {Massey}, {Masters}, {Matarrese}, {Matsuoka}, {Matthew}, {Maughan}, {Mauri}, {Maurin}, {Maurogordato}, {McCarthy}, {McConnachie}, {McCracken}, {McDonald},
  {McEwen}, {McPartland}, {Medinaceli}, {Mehta}, {Mei}, {Melchior}, {Melin}, {M{\'e}nard}, {Mendes}, {Mendez-Abreu}, {Meneghetti}, {Mercurio}, {Merlin}, {Metcalf}, {Meylan}, {Migliaccio}, {Mignoli}, {Miller}, {Miluzio}, {Milvang-Jensen}, {Mimoso}, {Miquel}, {Miyatake}, {Mobasher}, {Mohr}, {Monaco}, {Mongui{\'o}}, {Montoro}, {Mora}, {Moradinezhad Dizgah}, {Moresco}, {Moretti}, {Morgante}, {Morisset}, {Moriya}, {Morris}, {Mortlock}, {Moscardini}, {Mota}, {Moustakas}, {Moutard}, {M{\"u}ller}, {Munari}, {Murphree}, {Murray}, {Murray}, {Musi}, {Nadathur}, {Nagam}, {Nagao}, {Naidoo}, {Nakajima}, {Nally}, {Natoli}, {Navarro-Alsina}, {Navarro Girones}, {Neissner}, {Nersesian}, {Nesseris}, {Nguyen-Kim}, {Nicastro}, {Nichol}, {Nielbock}, {Niemi}, {Nieto}, {Nilsson}, {Noller}, {Norberg}, {Nourizonoz}, {Ntelis}, {Nucita}, {Nugent}, {Nunes}, {Nutma}, {Ocampo}, {Odier}, {Oesch}, {Oguri}, {Magalhaes Oliveira}, {Onoue}, {Oosterbroek}, {Oppizzi}, {Ordenovic}, {Osato}, {Pacaud}, {Pace}, {Padilla}, {Paech}, {Pagano}, {Page},
  {Palazzi}, {Paltani}, {Pamuk}, {Pandolfi}, {Paoletti}, {Paolillo}, {Papaderos}, {Pardede}, {Parimbelli}, {Parmar}, {Partmann}, {Pasian}, {Passalacqua}, {Paterson}, {Patrizii}, {Pattison}, {Paulino-Afonso}, {Paviot}, {Peacock}, {Pearce}, {Pedersen}, {Peel}, {Peletier}, {Pellejero Ibanez}, {Pello}, {Penny}, {Percival}, {Perez-Garrido}, {Perotto}, {Pettorino}, {Pezzotta}, {Pezzuto}, {Philippon}, {Piersanti}, {Pietroni}, {Piga}, {Pilo}, {Pires}, {Pisani}, {Pizzella}, {Pizzuti}, {Plana}, {Polenta}, {Pollack}, {Poncet}, {P{\"o}ntinen}, {Pool}, {Popa}, {Popa}, {Popp}, {Porciani}, {Porth}, {Potter}, {Poulain}, {Pourtsidou}, {Pozzetti}, {Prandoni}, {Pratt}, {Prezelus}, {Prieto}, {Pugno}, {Quai}, {Quilley}, {Racca}, {Raccanelli}, {R{\'a}cz}, {Radinovi{\'c}}, {Radovich}, {Ragagnin}, {Ragnit}, {Raison}, {Ramos-Chernenko}, {Ranc}, {Raylet}, {Rebolo}, {Refregier}, {Reimberg}, {Reiprich}, {Renk}, {Renzi}, {Retre}, {Revaz}, {Reyl{\'e}}, {Reynolds}, {Rhodes}, {Ricci}, {Ricci}, {Riccio}, {Ricken}, {Rissanen}, {Risso}, {Rix},
  {Robin}, {Rocca-Volmerange}, {Rocci}, {Rodenhuis}, {Rodighiero}, {Rodriguez Monroy}, {Rollins}, {Romanello}, {Roman}, {Romelli}, {Romero-Gomez}, {Roncarelli}, {Rosati}, {Rosset}, {Rossetti}, {Roster}, {Rottgering}, {Rozas-Fern{\'a}ndez}, {Ruane}, {Rubino-Martin}, {Rudolph}, {Ruppin}, {Rusholme}, {Sacquegna}, {S{\'a}ez-Casares}, {Saga}, {Saglia}, {Sahl{\'e}n}, {Saifollahi}, {Sakr}, {Salvalaggio}, {Salvaterra}, {Salvati}, {Salvato}, {Salvignol}, {S{\'a}nchez}, {Sanchez}, {Sanders}, {Sapone}, {Saponara}, {Sarpa}, {Sarron}, {Sartori}, {Sassolas}, {Sauniere}, {Sauvage}, {Sawicki}, {Scaramella}, {Scarlata}, {Scharr{\'e}}, {Schaye}, {Schewtschenko}, {Schindler}, {Schinnerer}, {Schirmer}, {Schmidt}, {Schmidt}, {Schmidt}, {Schneider}, {Schneider}, {Schneider}, {Sch{\"o}neberg}, {Schrabback}, {Schultheis}, {Schulz}, {Schwartz}, {Sciotti}, {Scodeggio}, {Scognamiglio}, {Scott}, {Scottez}, {Secroun}, {Sefusatti}, {Seidel}, {Seiffert}, {Sellentin}, {Selwood}, {Semboloni}, {Sereno}, {Serjeant}, {Serrano}, {Shankar},
  {Sharples}, {Short}, {Shulevski}, {Shuntov}, {Sias}, {Sikkema}, {Silvestri}, {Simon}, {Sirignano}, {Sirri}, {Skottfelt}, {Slezak}, {Sluse}, {Smith}, {Smith}, {Smith}, {Smit}, {Soldano}, {Solheim}, {Sorce}, {Sorrenti}, {Soubrie}, {Spinoglio}, {Spurio Mancini}, {Stadel}, {Stagnaro}, {Stanco}, {Stanford}, {Starck}, {Stassi}, {Steinwagner}, {Stern}, {Stone}, {Strada}, {Strafella}, {Stramaccioni}, {Surace}, {Sureau}, {Suyu}, {Swindells}, {Szafraniec}, {Szapudi}, {Taamoli}, {Talia}, {Tallada-Cresp{\'\i}}, {Tanidis}, {Tao}, {Tarr{\'\i}o}, {Tavagnacco}, {Taylor}, {Taylor}, {Taylor}, {Teixeira}, {Tenti}, {Teodoro Idiago}, {Teplitz}, {Tereno}, {Tessore}, {Testa}, {Testera}, {Tewes}, {Teyssier}, {Theret}, {Thizy}, {Thomas}, {Toba}, {Toft}, {Toledo-Moreo}, {Tolstoy}, {Tommasi}, {Torbaniuk}, {Torradeflot}, {Tortora}, {Tosi}, {Tosti}, {Trifoglio}, {Troja}, {Trombetti}, {Tronconi}, {Tsedrik}, {Tsyganov}, {Tucci}, {Tutusaus}, {Uhlemann}, {Ulivi}, {Urbano}, {Vacher}, {Vaillon}, {Valdes}, {Valentijn}, {Valenziano},
  {Valieri}, {Valiviita}, {Van den Broeck}, {Vassallo}, {Vavrek}, {Venemans}, {Venhola}, {Ventura}, {Verdoes Kleijn}, {Vergani}, {Verma}, {Vernizzi}, {Veropalumbo}, {Verza}, {Vescovi}, {Vibert}, {Viel}, {Vielzeuf}, {Viglione}, {Viitanen}, {Villaescusa-Navarro}, {Vinciguerra}, {Visticot}, {Voggel}, {von Wietersheim-Kramsta}, {Vriend}, {Wachter}, {Walmsley}, {Walth}, {Walton}, {Walton}, {Wander}, {Wang}, {Wang}, {Weaver}, {Weller}, {Whalen}, {Wiesmann}, {Wilde}, {Williams}, {Winther}, {Wittje}, {Wong}, {Wright}, {Yankelevich}, {Yeung}, {Youles}, {Yung}, {Zacchei}, {Zalesky}, {Zamorani}, {Zamorano Vitorelli}, {Zanoni Marc}, {Zennaro}, {Zerbi}, {Zinchenko}, {Zoubian}, {Zucca}, \& {Zumalacarregui}}]{Euclid}
{Euclid Collaboration}, {Mellier}, Y., {Abdurro'uf}, {et~al.} 2024, arXiv e-prints, arXiv:2405.13491, \dodoi{10.48550/arXiv.2405.13491}

\bibitem[{Feng {et~al.}(2024)Feng, Bouman, \& Freeman}]{Feng2024}
Feng, B.~T., Bouman, K.~L., \& Freeman, W.~T. 2024, The Astrophysical Journal, 975, 201, \dodoi{10.3847/1538-4357/ad737f}

\bibitem[{Feng {et~al.}(2023)Feng, Smith, Rubinstein, Chang, Bouman, \& Freeman}]{Feng2023}
Feng, B.~T., Smith, J., Rubinstein, M., {et~al.} 2023, 2023 IEEE/CVF International Conference on Computer Vision (ICCV), 10486.
\newblock \url{https://api.semanticscholar.org/CorpusID:258298618}

\bibitem[{{Fl{\"o}ss} {et~al.}(2024){Fl{\"o}ss}, {Coulton}, {Duivenvoorden}, {Villaescusa-Navarro}, \& {Wandelt}}]{Floss2024}
{Fl{\"o}ss}, T., {Coulton}, W.~R., {Duivenvoorden}, A.~J., {Villaescusa-Navarro}, F., \& {Wandelt}, B.~D. 2024, \mnras, 533, 423, \dodoi{10.1093/mnras/stae1818}

\bibitem[{Gelman(2008)}]{gelman2008objections}
Gelman, A. 2008, Bayesian Analysis, 3, 445

\bibitem[{Gelman {et~al.}(2013)Gelman, Carlin, Stern, Dunson, Vehtari, \& Rubin}]{BDA}
Gelman, A., Carlin, J.~B., Stern, H.~S., {et~al.} 2013, Bayesian data analysis, 3rd edn., Chapman \& Hall/CRC Texts in Statistical Science Series (Boca Raton, Florida: CRC).
\newblock \url{https://www.worldcat.org/title/bayesian-data-analysis/oclc/966614951?referer=br&ht=edition}

\bibitem[{{Genel} {et~al.}(2014){Genel}, {Vogelsberger}, {Springel}, {Sijacki}, {Nelson}, {Snyder}, {Rodriguez-Gomez}, {Torrey}, \& {Hernquist}}]{Generl2014-illustris}
{Genel}, S., {Vogelsberger}, M., {Springel}, V., {et~al.} 2014, Monthly Notices of the Royal Astronomical Society, 445, 175, \dodoi{10.1093/mnras/stu1654}

\bibitem[{{Ginsburg} {et~al.}(2019){Ginsburg}, {Sip{\H{o}}cz}, {Brasseur}, {Cowperthwaite}, {Craig}, {Deil}, {Guillochon}, {Guzman}, {Liedtke}, {Lian Lim}, {Lockhart}, {Mommert}, {Morris}, {Norman}, {Parikh}, {Persson}, {Robitaille}, {Segovia}, {Singer}, {Tollerud}, {de Val-Borro}, {Valtchanov}, {Woillez}, {Astroquery Collaboration}, \& {a subset of astropy Collaboration}}]{Ginsburg2019-Astroquery}
{Ginsburg}, A., {Sip{\H{o}}cz}, B.~M., {Brasseur}, C.~E., {et~al.} 2019, The Astronomical Journal, 157, 98, \dodoi{10.3847/1538-3881/aafc33}

\bibitem[{Graikos {et~al.}(2022)Graikos, Malkin, Jojic, \& Samaras}]{Graikos2022_plug_and_play}
Graikos, A., Malkin, N., Jojic, N., \& Samaras, D. 2022, in Advances in Neural Information Processing Systems, ed. S.~Koyejo, S.~Mohamed, A.~Agarwal, D.~Belgrave, K.~Cho, \& A.~Oh, Vol.~35 (Curran Associates, Inc.), 14715--14728.
\newblock \url{https://proceedings.neurips.cc/paper_files/paper/2022/file/5e6cec2a9520708381fe520246018e8b-Paper-Conference.pdf}

\bibitem[{Hartley(1958)}]{Hartley1958}
Hartley, H.~O. 1958, Biometrics, 14, 174.
\newblock \url{http://www.jstor.org/stable/2527783}

\bibitem[{{Heurtel-Depeiges} {et~al.}(2023){Heurtel-Depeiges}, {Burkhart}, {Ohana}, \& {R{\'e}galdo-Saint Blancard}}]{Heurtel-Depeiges2023}
{Heurtel-Depeiges}, D., {Burkhart}, B., {Ohana}, R., \& {R{\'e}galdo-Saint Blancard}, B. 2023, in Machine Learning and the Physical Sciences Workshop, NeurIPS 2023, arXiv:2310.16285.
\newblock \doarXiv{2310.16285}

\bibitem[{{Hezaveh} {et~al.}(2016){Hezaveh}, {Dalal}, {Holder}, {Kisner}, {Kuhlen}, \& {Perreault Levasseur}}]{Hezaveh2016}
{Hezaveh}, Y., {Dalal}, N., {Holder}, G., {et~al.} 2016, JCAP, 2016, 048, \dodoi{10.1088/1475-7516/2016/11/048}

\bibitem[{Ho {et~al.}(2020)Ho, Jain, \& Abbeel}]{Ho2020ddpm}
Ho, J., Jain, A., \& Abbeel, P. 2020, in Advances in Neural Information Processing Systems, ed. H.~Larochelle, M.~Ranzato, R.~Hadsell, M.~Balcan, \& H.~Lin, Vol.~33 (Curran Associates, Inc.), 6840--6851.
\newblock \url{https://proceedings.neurips.cc/paper_files/paper/2020/file/4c5bcfec8584af0d967f1ab10179ca4b-Paper.pdf}

\bibitem[{Hu {et~al.}(2022)Hu, yelong shen, Wallis, Allen-Zhu, Li, Wang, Wang, \& Chen}]{LORA}
Hu, E.~J., yelong shen, Wallis, P., {et~al.} 2022, in International Conference on Learning Representations.
\newblock \url{https://openreview.net/forum?id=nZeVKeeFYf9}

\bibitem[{Hyv{{\"a}}rinen(2005)}]{Hyvarinen2005}
Hyv{{\"a}}rinen, A. 2005, Journal of Machine Learning Research, 6, 695.
\newblock \url{http://jmlr.org/papers/v6/hyvarinen05a.html}

\bibitem[{Kadkhodaie {et~al.}(2024)Kadkhodaie, Guth, Simoncelli, \& Mallat}]{Kadkodaie2023}
Kadkhodaie, Z., Guth, F., Simoncelli, E.~P., \& Mallat, S. 2024, in The Twelfth International Conference on Learning Representations.
\newblock \url{https://openreview.net/forum?id=ANvmVS2Yr0}

\bibitem[{Karchev {et~al.}(2022)Karchev, Anau~Montel, Coogan, \& Weniger}]{Karchev2022}
Karchev, K., Anau~Montel, N., Coogan, A., \& Weniger, C. 2022, in Machine Learning and the Physical Sciences Workshop, NeurIPS 2022.
\newblock \doarXiv{2211.04365}

\bibitem[{Karras {et~al.}(2022)Karras, Aittala, Aila, \& Laine}]{Karras2022design}
Karras, T., Aittala, M., Aila, T., \& Laine, S. 2022, in Advances in Neural Information Processing Systems, ed. S.~Koyejo, S.~Mohamed, A.~Agarwal, D.~Belgrave, K.~Cho, \& A.~Oh, Vol.~35 (Curran Associates, Inc.), 26565--26577.
\newblock \url{https://proceedings.neurips.cc/paper_files/paper/2022/file/a98846e9d9cc01cfb87eb694d946ce6b-Paper-Conference.pdf}

\bibitem[{Kawar {et~al.}(2022)Kawar, Elad, Ermon, \& Song}]{Kawar2022inverse}
Kawar, B., Elad, M., Ermon, S., \& Song, J. 2022, in ICLR Workshop on Deep Generative Models for Highly Structured Data.
\newblock \url{https://openreview.net/forum?id=BExXihVOvWq}

\bibitem[{Kingma {et~al.}(2021)Kingma, Salimans, Poole, \& Ho}]{Kingma2021vdm}
Kingma, D., Salimans, T., Poole, B., \& Ho, J. 2021, in Advances in Neural Information Processing Systems, ed. M.~Ranzato, A.~Beygelzimer, Y.~Dauphin, P.~Liang, \& J.~W. Vaughan, Vol.~34 (Curran Associates, Inc.), 21696--21707.
\newblock \url{https://proceedings.neurips.cc/paper_files/paper/2021/file/b578f2a52a0229873fefc2a4b06377fa-Paper.pdf}

\bibitem[{Kingma \& Ba(2015)}]{Diederik2015-adam}
Kingma, D.~P., \& Ba, J. 2015, in 3rd International Conference on Learning Representations, {ICLR} 2015, San Diego, CA, USA, May 7-9, 2015, Conference Track Proceedings, ed. Y.~Bengio \& Y.~LeCun.
\newblock \url{http://arxiv.org/abs/1412.6980}

\bibitem[{Kingma \& Welling(2014)}]{Kingma2014}
Kingma, D.~P., \& Welling, M. 2014, in 2nd International Conference on Learning Representations, {ICLR} 2014, Banff, AB, Canada, April 14-16, 2014, Conference Track Proceedings.
\newblock \url{https://openreview.net/forum?id=33X9fd2-9FyZd}

\bibitem[{Kullback \& Leibler(1951)}]{KLdiv}
Kullback, S., \& Leibler, R.~A. 1951, The Annals of Mathematical Statistics, 22, 79 , \dodoi{10.1214/aoms/1177729694}

\bibitem[{Lanusse {et~al.}(2021)Lanusse, Mandelbaum, Ravanbakhsh, Li, Freeman, \& Póczos}]{lanusse_deep_2021}
Lanusse, F., Mandelbaum, R., Ravanbakhsh, S., {et~al.} 2021, Monthly Notices of the Royal Astronomical Society, 504, 5543, \dodoi{10.1093/mnras/stab1214}

\bibitem[{Lecun {et~al.}(1998)Lecun, Bottou, Bengio, \& Haffner}]{LeCun1998-mnist}
Lecun, Y., Bottou, L., Bengio, Y., \& Haffner, P. 1998, Proceedings of the IEEE, 86, 2278, \dodoi{10.1109/5.726791}

\bibitem[{Legin {et~al.}(2023)Legin, Ho, Lemos, Perreault-Levasseur, Ho, Hezaveh, \& Wandelt}]{ronan2024}
Legin, R., Ho, M., Lemos, P., {et~al.} 2023, Monthly Notices of the Royal Astronomical Society: Letters, 527, L173, \dodoi{10.1093/mnrasl/slad152}

\bibitem[{Lemos {et~al.}(2023)Lemos, Coogan, Hezaveh, \& Perreault-Levasseur}]{lemos2023-tarp}
Lemos, P., Coogan, A., Hezaveh, Y., \& Perreault-Levasseur, L. 2023, in Proceedings of Machine Learning Research, Vol. 202, Proceedings of the 40th International Conference on Machine Learning, ed. A.~Krause, E.~Brunskill, K.~Cho, B.~Engelhardt, S.~Sabato, \& J.~Scarlett (PMLR), 19256--19273.
\newblock \url{https://proceedings.mlr.press/v202/lemos23a.html}

\bibitem[{Lemos {et~al.}(2024)Lemos, Sharief, Malkin, Perreault-Levasseur, \& Hezaveh}]{lemos2024-pqmass}
Lemos, P., Sharief, S., Malkin, N., Perreault-Levasseur, L., \& Hezaveh, Y. 2024, {PQMass}: {Probabilistic} {Assessment} of the {Quality} of {Generative} {Models} using {Probability} {Mass} {Estimation},  arXiv, \dodoi{10.48550/arXiv.2402.04355}

\bibitem[{{Lintott} {et~al.}(2011){Lintott}, {Schawinski}, {Bamford}, {Slosar}, {Land}, {Thomas}, {Edmondson}, {Masters}, {Nichol}, {Raddick}, {Szalay}, {Andreescu}, {Murray}, \& {Vandenberg}}]{Lintott2011-GalaxyZoo}
{Lintott}, C., {Schawinski}, K., {Bamford}, S., {et~al.} 2011, Monthly Notices of the Royal Astronomical Society, 410, 166, \dodoi{10.1111/j.1365-2966.2010.17432.x}

\bibitem[{Little(2006)}]{little2006calibrated}
Little, R.~J. 2006, The American Statistician, 60, 213

\bibitem[{McLachlan \& Krishnan(2007)}]{mclachlan2007algorithm}
McLachlan, G., \& Krishnan, T. 2007, The EM Algorithm and Extensions, Wiley Series in Probability and Statistics (Wiley).
\newblock \url{https://books.google.ca/books?id=NBawzaWoWa8C}

\bibitem[{Mudur {et~al.}(2024)Mudur, Cuesta-Lazaro, \& Finkbeiner}]{Mudur2024}
Mudur, N., Cuesta-Lazaro, C., \& Finkbeiner, D.~P. 2024, The Astrophysical Journal, 978, 64, \dodoi{10.3847/1538-4357/ad8bc3}

\bibitem[{Mudur \& Finkbeiner(2022)}]{Mudur2022}
Mudur, N., \& Finkbeiner, D.~P. 2022, in Machine Learning and the Physical Sciences Workshop, NeurIPS 2022.
\newblock \doarXiv{2211.12444}

\bibitem[{Murata {et~al.}(2023)Murata, Saito, Lai, Takida, Uesaka, Mitsufuji, \& Ermon}]{murata_gibbsddrm_2023}
Murata, N., Saito, K., Lai, C.-H., {et~al.} 2023, in Proceedings of Machine Learning Research, Vol. 202, Proceedings of the 40th International Conference on Machine Learning, ed. A.~Krause, E.~Brunskill, K.~Cho, B.~Engelhardt, S.~Sabato, \& J.~Scarlett (PMLR), 25501--25522.
\newblock \url{https://proceedings.mlr.press/v202/murata23a.html}

\bibitem[{{Nelson} {et~al.}(2019){Nelson}, {Springel}, {Pillepich}, {Rodriguez-Gomez}, {Torrey}, {Genel}, {Vogelsberger}, {Pakmor}, {Marinacci}, {Weinberger}, {Kelley}, {Lovell}, {Diemer}, \& {Hernquist}}]{Nelson2019}
{Nelson}, D., {Springel}, V., {Pillepich}, A., {et~al.} 2019, Computational Astrophysics and Cosmology, 6, 2, \dodoi{10.1186/s40668-019-0028-x}

\bibitem[{Nichol \& Dhariwal(2021)}]{Nichol2021improved}
Nichol, A.~Q., \& Dhariwal, P. 2021, in Proceedings of Machine Learning Research, Vol. 139, Proceedings of the 38th International Conference on Machine Learning, ed. M.~Meila \& T.~Zhang (PMLR), 8162--8171.
\newblock \url{https://proceedings.mlr.press/v139/nichol21a.html}

\bibitem[{Ono {et~al.}(2024)Ono, Park, Mudur, Ni, Cuesta-Lazaro, \& Villaescusa-Navarro}]{Ono2024}
Ono, V., Park, C.~F., Mudur, N., {et~al.} 2024, The Astrophysical Journal, 970, 174, \dodoi{10.3847/1538-4357/ad5957}

\bibitem[{{Peng} {et~al.}(2006){Peng}, {Impey}, {Rix}, {Kochanek}, {Keeton}, {Falco}, {Leh{\'a}r}, \& {McLeod}}]{Peng2006}
{Peng}, C.~Y., {Impey}, C.~D., {Rix}, H.-W., {et~al.} 2006, ApJ, 649, 616, \dodoi{10.1086/506266}

\bibitem[{{Remy} {et~al.}(2023){Remy}, {Lanusse}, {Jeffrey}, {Liu}, {Starck}, {Osato}, \& {Schrabback}}]{Remy2023}
{Remy}, B., {Lanusse}, F., {Jeffrey}, N., {et~al.} 2023, A\&A, 672, A51, \dodoi{10.1051/0004-6361/202243054}

\bibitem[{Rezende \& Mohamed(2015)}]{Rezende2015_NF}
Rezende, D., \& Mohamed, S. 2015, in Proceedings of Machine Learning Research, Vol.~37, Proceedings of the 32nd International Conference on Machine Learning, ed. F.~Bach \& D.~Blei (Lille, France: PMLR), 1530--1538.
\newblock \url{https://proceedings.mlr.press/v37/rezende15.html}

\bibitem[{Ronneberger {et~al.}(2015)Ronneberger, Fischer, \& Brox}]{Ronnenberger2015}
Ronneberger, O., Fischer, P., \& Brox, T. 2015, in Medical Image Computing and Computer-Assisted Intervention -- MICCAI 2015, ed. N.~Navab, J.~Hornegger, W.~M. Wells, \& A.~F. Frangi (Cham: Springer International Publishing), 234--241, \dodoi{10.1007/978-3-319-24574-4_28}

\bibitem[{{Rose} {et~al.}(2024){Rose}, {Torrey}, {Villaescusa-Navarro}, {Lisanti}, {Nguyen}, {Roy}, {Kollmann}, {Vogelsberger}, {Cyr-Racine}, {Medvedev}, {Genel}, {Angl{\'e}s-Alc{\'a}zar}, {Kallivayalil}, {Wang}, {Costanza}, {O'Neil}, {Roche}, {Karmakar}, {Garcia}, {Low}, {Lin}, {Mostow}, {Cruz}, {Caputo}, {Farahi}, {Mu{\~n}oz}, {Necib}, {Teyssier}, {Dalcanton}, \& {Spergel}}]{Rose2024-DREAMS}
{Rose}, J.~C., {Torrey}, P., {Villaescusa-Navarro}, F., {et~al.} 2024, arXiv e-prints, arXiv:2405.00766, \dodoi{10.48550/arXiv.2405.00766}

\bibitem[{Rozet {et~al.}(2024)Rozet, Andry, Lanusse, \& Louppe}]{Rozet2024}
Rozet, F., Andry, G., Lanusse, F., \& Louppe, G. 2024, in The Thirty-eighth Annual Conference on Neural Information Processing Systems.
\newblock \url{https://openreview.net/forum?id=7v88Fh6iSM}

\bibitem[{Rozet \& Louppe(2023)}]{Rozet2023}
Rozet, F., \& Louppe, G. 2023, in Advances in Neural Information Processing Systems, ed. A.~Oh, T.~Naumann, A.~Globerson, K.~Saenko, M.~Hardt, \& S.~Levine, Vol.~36 (Curran Associates, Inc.), 40521--40541.
\newblock \url{https://proceedings.neurips.cc/paper_files/paper/2023/file/7f7fa581cc8a1970a4332920cdf87395-Paper-Conference.pdf}

\bibitem[{Rupp {et~al.}(2004)Rupp, Dey, \& Zumbo}]{rupp2004bayes}
Rupp, A.~A., Dey, D.~K., \& Zumbo, B.~D. 2004, Structural Equation Modeling, 11, 424

\bibitem[{{Ruth}(2024)}]{RuthWilliam2024}
{Ruth}, W. 2024, arXiv e-prints, arXiv:2401.00945, \dodoi{10.48550/arXiv.2401.00945}

\bibitem[{{Smith} {et~al.}(2022){Smith}, {Geach}, {Jackson}, {Arora}, {Stone}, \& {Courteau}}]{Smith2022}
{Smith}, M.~J., {Geach}, J.~E., {Jackson}, R.~A., {et~al.} 2022, Monthly Notices of the Royal Astronomical Society, 511, 1808, \dodoi{10.1093/mnras/stac130}

\bibitem[{Sohl-Dickstein {et~al.}(2015)Sohl-Dickstein, Weiss, Maheswaranathan, \& Ganguli}]{Sohl-Dickstein2015diffusion}
Sohl-Dickstein, J., Weiss, E., Maheswaranathan, N., \& Ganguli, S. 2015, in Proceedings of Machine Learning Research, Vol.~37, Proceedings of the 32nd International Conference on Machine Learning, ed. F.~Bach \& D.~Blei (Lille, France: PMLR), 2256--2265.
\newblock \url{https://proceedings.mlr.press/v37/sohl-dickstein15.html}

\bibitem[{Song {et~al.}(2021{\natexlab{a}})Song, Durkan, Murray, \& Ermon}]{Song2021MLE}
Song, Y., Durkan, C., Murray, I., \& Ermon, S. 2021{\natexlab{a}}, in Advances in Neural Information Processing Systems, ed. M.~Ranzato, A.~Beygelzimer, Y.~Dauphin, P.~Liang, \& J.~W. Vaughan, Vol.~34 (Curran Associates, Inc.), 1415--1428.
\newblock \url{https://proceedings.neurips.cc/paper_files/paper/2021/file/0a9fdbb17feb6ccb7ec405cfb85222c4-Paper.pdf}

\bibitem[{Song \& Ermon(2019)}]{Song2019}
Song, Y., \& Ermon, S. 2019, in Advances in Neural Information Processing Systems, ed. H.~Wallach, H.~Larochelle, A.~Beygelzimer, F.~d\textquotesingle Alch\'{e}-Buc, E.~Fox, \& R.~Garnett, Vol.~32 (Curran Associates, Inc.).
\newblock \url{https://proceedings.neurips.cc/paper_files/paper/2019/file/3001ef257407d5a371a96dcd947c7d93-Paper.pdf}

\bibitem[{Song \& Ermon(2020)}]{Song2020improved}
Song, Y., \& Ermon, S. 2020, in Advances in Neural Information Processing Systems, ed. H.~Larochelle, M.~Ranzato, R.~Hadsell, M.~Balcan, \& H.~Lin, Vol.~33 (Curran Associates, Inc.), 12438--12448.
\newblock \url{https://proceedings.neurips.cc/paper_files/paper/2020/file/92c3b916311a5517d9290576e3ea37ad-Paper.pdf}

\bibitem[{Song {et~al.}(2022)Song, Shen, Xing, \& Ermon}]{Song2022mri}
Song, Y., Shen, L., Xing, L., \& Ermon, S. 2022, in The Tenth International Conference on Learning Representations.
\newblock \url{https://openreview.net/forum?id=vaRCHVj0uGI}

\bibitem[{Song {et~al.}(2021{\natexlab{b}})Song, Sohl{-}Dickstein, Kingma, Kumar, Ermon, \& Poole}]{Song2021sde}
Song, Y., Sohl{-}Dickstein, J., Kingma, D.~P., {et~al.} 2021{\natexlab{b}}, in 9th International Conference on Learning Representations, {ICLR} 2021, Virtual Event, Austria, May 3-7, 2021 (OpenReview.net).
\newblock \url{https://openreview.net/forum?id=PxTIG12RRHS}

\bibitem[{Stone {et~al.}(2024)Stone, Adam, Coogan, Yantovski-Barth, Filipp, Setiawan, Core, Legin, Wilson, Barco, Hezaveh, \& Perreault-Levasseur}]{Stone2024}
Stone, C., Adam, A., Coogan, A., {et~al.} 2024, Journal of Open Source Software, 9, 7081, \dodoi{10.21105/joss.07081}

\bibitem[{{Stone} {et~al.}(2023){Stone}, {Courteau}, {Cuillandre}, {Hezaveh}, {Perreault-Levasseur}, \& {Arora}}]{Stone2023}
{Stone}, C.~J., {Courteau}, S., {Cuillandre}, J.-C., {et~al.} 2023, MNRAS, \dodoi{10.1093/mnras/stad2477}

\bibitem[{Sun \& Bouman(2021)}]{Sun2020}
Sun, H., \& Bouman, K.~L. 2021, Proceedings of the AAAI Conference on Artificial Intelligence, 35, 2628, \dodoi{10.1609/aaai.v35i3.16366}

\bibitem[{van~de Schoot {et~al.}(2021)van~de Schoot, Depaoli, King, Kramer, Märtens, Tadesse, Vannucci, Gelman, Veen, Willemsen, \& Yau}]{van_de_schoot_bayesian_2021}
van~de Schoot, R., Depaoli, S., King, R., {et~al.} 2021, Nature Reviews Methods Primers, 1, 1, \dodoi{10.1038/s43586-020-00001-2}

\bibitem[{Vandegar {et~al.}(2021)Vandegar, Kagan, Wehenkel, \& Louppe}]{vandegar2021}
Vandegar, M., Kagan, M., Wehenkel, A., \& Louppe, G. 2021, in Proceedings of Machine Learning Research, Vol. 130, Proceedings of The 24th International Conference on Artificial Intelligence and Statistics, ed. A.~Banerjee \& K.~Fukumizu (PMLR), 2107--2115.
\newblock \url{https://proceedings.mlr.press/v130/vandegar21a.html}

\bibitem[{{Vegetti} \& {Vogelsberger}(2014)}]{Vegetti2014}
{Vegetti}, S., \& {Vogelsberger}, M. 2014, MNRAS, 442, 3598, \dodoi{10.1093/mnras/stu1284}

\bibitem[{Vetter {et~al.}(2024)Vetter, Moss, Schr{\"o}der, Gao, \& Macke}]{vetter2024-sourcerer}
Vetter, J., Moss, G., Schr{\"o}der, C., Gao, R., \& Macke, J.~H. 2024, in The Thirty-eighth Annual Conference on Neural Information Processing Systems.
\newblock \url{https://openreview.net/forum?id=0cgDDa4OFr}

\bibitem[{Villaescusa-Navarro {et~al.}(2022)Villaescusa-Navarro, Genel, Angl{\'{e} }s-Alc{\'{a}}zar, Thiele, Dave, Narayanan, Nicola, Li, Villanueva-Domingo, Wandelt, Spergel, Somerville, Matilla, Mohammad, Hassan, Shao, Wadekar, Eickenberg, Wong, Contardo, Jo, Moser, Lau, Valle, Perez, Nagai, Battaglia, \& Vogelsberger}]{CMD}
Villaescusa-Navarro, F., Genel, S., Angl{\'{e} }s-Alc{\'{a}}zar, D., {et~al.} 2022, The Astrophysical Journal Supplement Series, 259, 61, \dodoi{10.3847/1538-4365/ac5ab0}

\bibitem[{Vincent(2011)}]{Vincent2011}
Vincent, P. 2011, Neural Comput., 23, 1661, \dodoi{10.1162/NECO\_a\_00142}

\bibitem[{{Welch} {et~al.}(2022){Welch}, {Coe}, {Zackrisson}, {de Mink}, {Ravindranath}, {Anderson}, {Brammer}, {Bradley}, {Yoon}, {Kelly}, {Diego}, {Windhorst}, {Zitrin}, {Dimauro}, {Jim{\'e}nez-Teja}, {Abdurro'uf}, {Nonino}, {Acebron}, {Andrade-Santos}, {Avila}, {Bayliss}, {Ben{\'\i}tez}, {Broadhurst}, {Bhatawdekar}, {Brada{\v{c}}}, {Caminha}, {Chen}, {Eldridge}, {Farag}, {Florian}, {Frye}, {Fujimoto}, {Gomez}, {Henry}, {Hsiao}, {Hutchison}, {James}, {Joyce}, {Jung}, {Khullar}, {Larson}, {Mahler}, {Mandelker}, {McCandliss}, {Morishita}, {Newshore}, {Norman}, {O'Connor}, {Oesch}, {Oguri}, {Ouchi}, {Postman}, {Rigby}, {Ryan}, {Sharma}, {Sharon}, {Strait}, {Strolger}, {Timmes}, {Toft}, {Trenti}, {Vanzella}, \& {Vikaeus}}]{Welch2022}
{Welch}, B., {Coe}, D., {Zackrisson}, E., {et~al.} 2022, ApJL, 940, L1, \dodoi{10.3847/2041-8213/ac9d39}

\bibitem[{{Wong} {et~al.}(2020){Wong}, {Suyu}, {Chen}, {Rusu}, {Million}, {Sluse}, {Bonvin}, {Fassnacht}, {Taubenberger}, {Auger}, {Birrer}, {Chan}, {Courbin}, {Hilbert}, {Tihhonova}, {Treu}, {Agnello}, {Ding}, {Jee}, {Komatsu}, {Shajib}, {Sonnenfeld}, {Blandford}, {Koopmans}, {Marshall}, \& {Meylan}}]{holycow}
{Wong}, K.~C., {Suyu}, S.~H., {Chen}, G. C.~F., {et~al.} 2020, MNRAS, 498, 1420, \dodoi{10.1093/mnras/stz3094}

\bibitem[{Wu(1983)}]{JeffWu1983}
Wu, C. F.~J. 1983, The Annals of Statistics, 11, 95 , \dodoi{10.1214/aos/1176346060}

\bibitem[{{Xue} {et~al.}(2023){Xue}, {Li}, {Patel}, \& {Regier}}]{Xue2023}
{Xue}, Z., {Li}, Y., {Patel}, Y., \& {Regier}, J. 2023, in Workshop on Machine Learning for Astrophysics, ICML 2023, arXiv:2307.11122, \dodoi{10.48550/arXiv.2307.11122}

\end{thebibliography}

\newpage
\appendix

\section{Detailed proof of ascent property}
\label{app:proof}

There exists a long literature on the convergence properties of the generalized expectation maximization algorithm~\citep{Dempster1977,JeffWu1983, RuthWilliam2024}. We wish to show that the procedure outlined in \Alg{alg} incrementally increases the log-likelihood of observations, that is, it leads to successive priors models that allow us to incrementally increase the expected log-evidence of data for our fixed physical and noise models. More specifically,
\begin{equation}
\label{eq:inequality}
    \int d\mathbf{y} p(\mathbf{y}) \log p_{\theta_{\alpha+1}}(\mathbf{y})\geq \int d\mathbf{y} p(\mathbf{y}) \log p_{\theta_{\alpha}}(\mathbf{y}) \, ,
\end{equation}
where $ p_{\theta_{\alpha}}(\mathbf{y})= \int d\mathbf{x} p (\mathbf{y} \mid \mathbf{x}) p_{\theta_{\alpha}}(\mathbf{x})$ for every $\alpha$. We have:
\begin{align}
    & \int d\mathbf{y} p(\mathbf{y}) \left[ \log p_{\theta_{\alpha+1}}(\mathbf{y}) - \log p_{\theta_{\alpha}}(\mathbf{y})\right] \\
    & = \int d\mathbf{y} p(\mathbf{y})\log \left[\frac{  p_{\theta_{\alpha+1}}(\mathbf{y})}{ p_{\theta_{\alpha}}(\mathbf{y})}\right] \\ 
    & = \int d\mathbf{y} p(\mathbf{y}) \log \left[ \frac{\int d\mathbf{x} p_{\theta_{\alpha+1}}(\mathbf{x}) p(\mathbf{y}\mid \mathbf{x})}{p_{\theta_{\alpha}}(\mathbf{y})}\right]
    \label{eq:12}\\
     & = \int d\mathbf{y} p(\mathbf{y}) \log \left[\int d\mathbf{x} p_{\theta_{\alpha}}(\mathbf{x}\mid \mathbf{y})\frac{ p_{\theta_{\alpha+1}}(\mathbf{x}) }{p_{\theta_{\alpha}}(\mathbf{x})}\right]
   \label{eq:15} \\
    \label{eq:jensen}&\geq \int\int  d\mathbf{y} d \mathbf{x} p(\mathbf{y}) p_{\theta_\alpha} (\mathbf{x} \mid \mathbf{y}) \left[  \log p_{\theta_{\alpha+1}}(\mathbf{x}) - \log p_{\theta_{\alpha}}(\mathbf{x})\right]
\end{align}

Here, to go from line~(\ref{eq:12}) to line~(\ref{eq:15}), we have multiplied by $1={p_{\alpha}(\mathbf{x}\mid \mathbf{y})}/{p_{\alpha}(\mathbf{x}\mid \mathbf{y})}$ inside the $\mathbf{x}$ integral and used that the likelihood $p(\mathbf{y}\mid \mathbf{x})$ is the same for all $\alpha$'s, and have used that line~(\ref{eq:jensen}) follows from Jensen's inequality. Now, since the way we define $\theta_{\alpha+1}$ in our iterative update is by finding the values of $\theta$ that maximize the first terms in Eqn.~(\ref{eq:jensen}), and that, at worse, we could have $\theta_{\alpha+1}=\theta_{\alpha}$, we conclude that Eqn.~(\ref{eq:inequality}) follows. 

\section{Stationary Distribution}
\label{app:proof_stationary}

Under Definition \ref{def:mixture_prior}, in the large data limit, the next prior \(p_{\theta_{\alpha+1}}(\mathbf{x})\) is the expected posterior under prior \(p_{\theta_{\alpha}}(\mathbf{x})\) with observations from \(\mathbf{y} \sim p(\mathbf{y})\):
\begin{equation}
p_{\theta_{\alpha+1}}(\mathbf{x}) = \mathbb{E}_{\mathbf{y} \sim p(\mathbf{y})}[p_{\theta_\alpha}(\mathbf{x} \mid \mathbf{y} )]
\end{equation}
Samples from the observation distribution can be obtained by sampling the underlying population distribution $\mathbf{x} \sim p_{\theta^\star}(\mathbf{x})$ and using the forward process to calculate \(\mathbf{y} = A\mathbf{x} + \boldsymbol{\eta}\). The distribution of these samples is given by:
\begin{equation}
    p(\mathbf{y}) = \int p(\mathbf{y} \mid \mathbf{x}) p_{\theta^\star}(\mathbf{x}) \, d\mathbf{x}
\end{equation}
We observe that there is no change in the update if \(p_{\theta_\alpha}(\mathbf{x})\) has already converged to the stationary distribution, \(p_{\hat\theta}(\mathbf{x})\), defined in Eqn. \ref{eqn:thetastar} which has  marginal likelihood, or evidence, equal to the underlying population distribution, $ p_{\hat\theta}(\mathbf{y}) =  p(\mathbf{y})$:
\begin{align}
    p_{\theta_{\alpha + 1}}(\mathbf{x}) & = \mathbb{E}_{ \mathbf{y} \sim p(\mathbf{y})}\left[\frac{p(\mathbf{y} \mid \mathbf{x}) p_{\theta_\alpha}(\mathbf{x})}{p_{\theta_\alpha}(\mathbf{y})}\right] \\
    & = p_{\theta_\alpha}(\mathbf{x}) \int p(\mathbf{y} \mid \mathbf{x}) \frac{p(\mathbf{y})}{p_{\theta_\alpha}(\mathbf{y})} \, d\mathbf{y} \\
    & = p_{\theta_\alpha}(\mathbf{x})
\end{align}
Because if $p_{\theta_\alpha}(\mathbf{x})$ has converged to the stationary distribution $p_{\hat\theta}(\mathbf{x})$, then 
\begin{align}
    p_{\theta_\alpha}(\mathbf{y}) &= \int d\mathbf{x} \, p(\mathbf{y} \mid \mathbf{x}) p_{\theta_\alpha}(\mathbf{x})  \\ 
    &= \int d\mathbf{x} \, p(\mathbf{y} \mid \mathbf{x}) p_{\hat\theta}(\mathbf{x}) \\
    & = p_{\hat\theta}(\mathbf{y})\, .
\end{align}
However, in practice, we have finite samples, and we only approximate \(\mathbb{E}_{y \sim p_{\theta^\star}(\mathbf{y})}[p_{\theta_\alpha}(\mathbf{x} \mid \mathbf{y} )]\) at each iteration after training. For convergence to a distribution, we would need to have \(p_{\theta_\alpha}(\mathbf{y}) \approx p_{\theta^\star}(\mathbf{y})\). We also note that the prior distribution found after iterating does not necessarily converge to the underlying prior $p_{\theta^*}(\mathbf{x})$, but it is such that it has equal marginal likelihood, and therefore equally explains the observed data, making it a plausible hyper distribution.

\section{MNIST Experiment PQMass values}
\label{app:details_mnist}

\begin{figure}[tb]
    \centering
    \includegraphics[width = 3in]{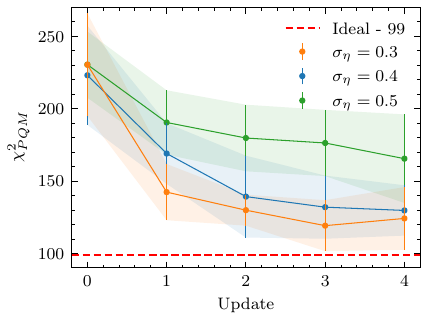}
    \caption{$\chi^2_{PQM}$ statistic to compare both distributions given samples, for the three experiments with MNIST, with different levels of observational noise.}
    \label{fig:pqm_mnist}
\end{figure}

Similar to the experiment with galaxies, we compute $\chi^2_{PQM}$ for the different levels of observational noise during the updates, as shown in \Fig{pqm_mnist}. In this case, we observe that the initial prior is closer to the true prior, given that there are only mode mismatches and not a distribution shift. Also, the difference after the updates between different observational noise levels $\sigma_{\boldsymbol{\eta}}$ is more evident.

\section{Models Architecture and training}
\label{app:model_architecture}

Galaxy images have a high dynamic range, featuring extremely bright cores and very faint structures. Therefore, we use the VP SDE for the galaxy experiment, as it better preserves faint, diffuse features and effectively handles large-scale brightness variations. For MNIST, we use the VE SDE, since the data is normalized to the $[0, 1]$ range, and VE excels at capturing the sharp, high-contrast features typical of handwritten digits.

Regarding the noise scales, as discussed by \citet{Song2020improved}, it is recommended to set $\sigma_{\text{min}} \ll 1$, and $\sigma_{\text{max}}$ to match the scale of the maximum Euclidean distance between any two points in the dataset. For MNIST, this distance is approximately $16$. We choose $\sigma_{\text{min}} = 10^{-5}$ and $\sigma_{\text{max}} = 100$. For galaxies, using VP, we follow \citet{Dia2023}, whose work also involves training an SBM on the SKIRT dataset, and select $\beta_{\text{min}} = 10^{-2}$ and $\beta_{\text{max}} = 20$.
The value for $\beta_{\text{max}}$ corresponds to scaling all values down by approximately five e-folds while maintaining constant variance~\citep[eq. 33]{Song2021sde}.

All models used for the galaxy experiments share the same architecture and training hyperparameters. The same applies to the MNIST models independently. In this appendix, we provide the necessary details to reproduce the experiments.

For the galaxy experiment, the architecture parameters within the \texttt{score-models} package are:

\begin{verbatim}
    "channels": 3,
    "nf": 64,
    "ch_mult": [1, 2, 2, 2],
    "num_res_blocks": 2
\end{verbatim}

And for MNIST:
\begin{verbatim}
    "channels": 1,
    "nf": 64,
    "ch_mult": [2, 2, 2],
    "num_res_blocks": 3
\end{verbatim}

We also use the \texttt{score-models} package to train the models. We use the Adam optimizer \citep{Diederik2015-adam}. For the galaxy experiments, we have a learning rate of $1 \times 10^{-4}$, a batch size of $256$, and an \texttt{ema\_decay} of $0.999$. For the MNIST experiments, we use a learning rate of $5 \times 10^{-5}$, a batch size of $256$, and an \texttt{ema\_decay} of $0.99$. For all experiments, we train for approximately $2.5\times10^5$ optimization steps. All unspecified hyperparameters are set to the default values of the \texttt{score-models} package.

We determined these configurations by testing 5 different parameter sets. In terms of compute resources, we performed training and inference (both prior and posterior sampling) on A100 GPUs. Each SBM model's training and sampling routine was conducted on a single A100 GPU. The MNIST models required approximately $14$ hours of training (wall-time), with $16$ GB of VRAM allocated. The galaxy models required about $20$ hours of training (wall-time), with $32$ GB of VRAM allocated, while posterior/prior sampling of a set of $1\,024$ samples took roughly $2$ hours (wall-time) and used $32$ GB of VRAM.

In total, for all experiments, we trained $12$ SBMs for MNIST, $30$ SBMs for galaxies, and conducted approximately $700$ rounds of prior sampling (to simulate observations) and posterior sampling (to create the training dataset for the next SBM prior), with $1\,024$ samples per set.

\section{Going from spiral to elliptical prior}
\label{app:reverse_experiment}

We conducted an additional experiment with the galaxy datasets, using the SBM trained on the elliptical dataset as the stationary distribution $p_{\theta}^\star (\mathbf{x})$ and the SBM trained on the spiral dataset as the initial prior $p_{\theta_0}(\mathbf{x})$. We performed $M = 6$ iterations using the same configuration as the experiment with the elliptical initial prior, with observational noise during the updates of $\sigma_{\boldsymbol{\eta}} = 1$. Although $\sigma_{\boldsymbol{\eta}}$ is the same for both experiments, the signal-to-noise ratio (SNR) for these observations is lower, since elliptical galaxies are dimmer than spiral ones. This effectively reduces the information in each observation. We hypothesize that this situation makes progress towards the stationary distribution more challenging and slower.

\begin{figure}[tb]
    \centering
    \includegraphics[width = 3in]{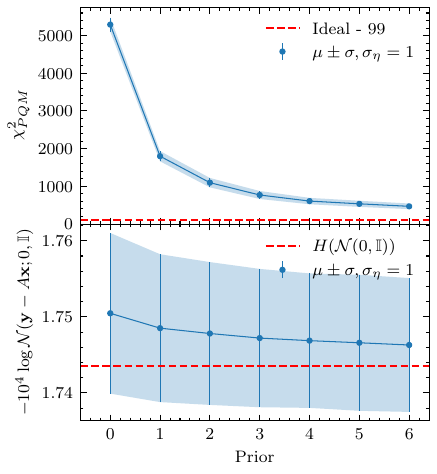}
    \caption{\textbf{Top:} $\chi^2_{PQM}$ statistic for the experiment with initial spiral galaxy prior and target elliptical prior. \textbf{Bottom:} Mean log-likelihood of the residuals $\mathbf{y} - A\mathbf{x}$ of $10\,240$ pairs $(\mathbf{y}, \mathbf{x} \sim p_{\theta_i}(\mathbf{x} \mid \mathbf{y}))$ at each iteration.}
    \label{fig:pqmass_reverse}
\end{figure}

\begin{figure*}[tb]
    \centering
    \includegraphics[width=6.4in]{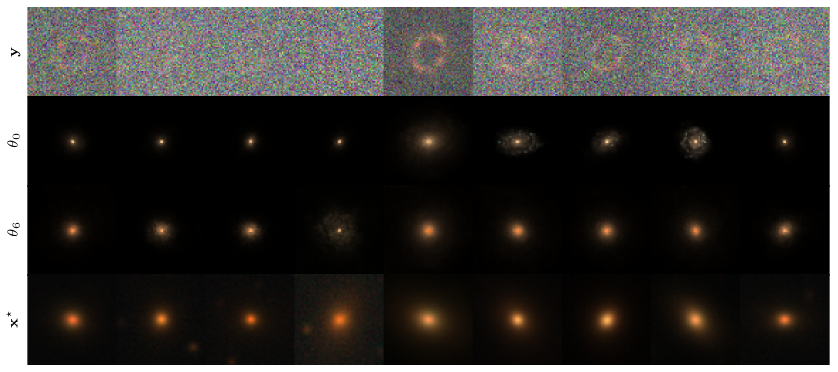}
    \caption{Posterior samples for the initial prior (\textbf{second row}) and the last update (\textbf{second row}) using the elliptical dataset as stationary distribution (\textbf{bottom row}). The noise level in the observations (\textbf{first row}) is $\sigma_{\boldsymbol{\eta}} = 2$.}
    \label{fig:main_reverse}
\end{figure*}

In \Fig{pqmass_reverse}, we observe progress in both the likelihood of the residuals and the value of $\chi^2_{PQM}$, similar to the other experiments. When using the initial and final priors $p_{\theta_i}$ to obtain posterior samples given observations from the stationary distribution, as observed in \Fig{main_reverse}, we note improvement in the reconstructions.

\newpage

\section{Galaxy Prior Samples}

\begin{figure*}[htb]
    \centering
    \includegraphics[width = 6.4in]{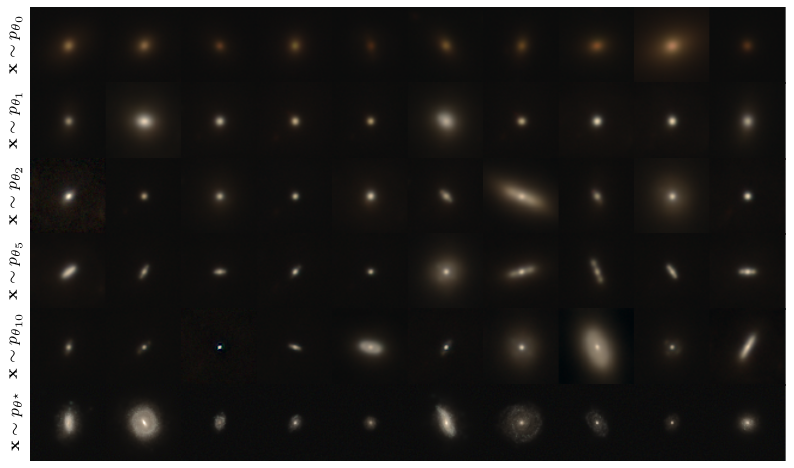}
    \caption{Prior samples from each model learned in the galaxy experiment with observational noise of $\sigma_{\boldsymbol{\eta}} = 1$.}
    \label{fig:prior_samples}
\end{figure*}

Finally, we can also observe the changes in $p_{\theta_i}(\mathbf{x})$ by generating samples at each iteration, as shown in \Fig{prior_samples}. For visualization purposes, we use the transformation $\mathbf{\bar{x}} = \log(\mathbf{x} - \mathbf{x}.\text{min}() + 1)$ and normalize it between $0$ and $1$, but now we use a fixed $x_0$ instead of $\mathbf{x}.\text{min}()$, and the same normalization values for all observations to observe changes in total flux between samples. In the first row, we observe samples from the initial prior, which appear dimmer and blurrier than samples from any subsequent prior. The final row contains samples from the target stationary distribution.

\end{document}